\documentstyle[psfig]{ar}

\begin{document}

\title{NUCLEAR SPIN CONVERSION IN POLYATOMIC MOLECULES}
\markboth{Chapovsky \& Hermans}{Nuclear Spin Conversion}

\author{P.L.~Chapovsky\affiliation{Institute of Automation and 
Electrometry, Russian Academy of Sciences,\\ 
630090 Novosibirsk, Russia;\\
        E-mail: chapovsky@iae.nsk.su}        
       L.J.F.~Hermans\affiliation{Huygens Laboratory, Leiden University,
P.O. Box 9504, 2300 RA Leiden, The Netherlands;\\
E-mail: hermans@molphys.leidenuniv.nl}}


%\date{\today}

\begin{abstract}
Except for ortho- and para-H$_2$, very little is known about nuclear 
spin isomers (or spin modifications) of molecules. The main reason 
is the lack of practical enrichment techniques. Recently a few 
enrichment methods were developed, which opened up new possibilities 
in the field. These methods are briefly reviewed. Substantial 
progress in the field has been made by the introduction of 
Light-Induced Drift as a gas-phase separation tool. This is illustrated 
by extensive data on CH$_3$F, which reveal that the gas-phase 
ortho-para conversion is governed by intramolecular mixing of the 
nuclear spin states. The role of ``direct'' ortho-para transitions 
is shown to be small. Various aspects of the conversion were 
investigated in detail: pressure and collision partner dependence, 
isotope effect, temperature dependence. The most decisive 
information on the spin conversion mechanism is derived from the 
observation of level-crossing resonances in an electric field and 
the Quantum Zeno effect induced by collisions. 
\end{abstract}

\maketitle

\section{INTRODUCTION}

\subsection{Hydrogen spin isomers}

At the beginning of the century physicists were puzzled by two
seemingly unrelated phenomena: the anomalous specific heat of
hydrogen discovered by Eucken in 1912 \cite{Eucken12} and the line 
intensity alternation in molecular spectra discovered by Mecke in 1925 
\cite{Mecke25}. The solution of these two problems was a landmark in 
the foundation of quantum mechanics. As was pointed out by Farkas, 
``it was a real triumph for theory when in 1929 Bonhoeffer and Harteck 
\cite{Bonhoeffer29} succeeded in bringing forward experimental evidence 
for the existence of the two different modifications of hydrogen'' 
\cite{Farkas35}.

Three quarters of a century later, the explanation of these problems
looks almost trivial. There are two nuclear spin isomers of H$_{2}$
which differ by total spin ($I$) of the two hydrogen nuclei: $I=1$ 
for ortho hydrogen and $I=0$ for para hydrogen. The symmetry of the 
molecular wave function upon interchange of the two protons allows 
only odd values of rotational angular momentum ($J$) for ortho and 
only even values of $J$ for para. The spin isomers of H$_{2}$ are 
extremely stable, having a conversion time on the order of 1~year 
at room temperature and 1~atm for pure hydrogen \cite{Farkas35}.

It is useful to recall the principle of the hydrogen spin isomer 
separation. A unique property of hydrogen molecules is their 
anomalously big rotational level spacing. The energy gap between 
the $J$=0 and $J$=1 states (first para and ortho states of hydrogen, 
respectively) is $\simeq 170$~K. This energy gap is much larger than 
the boiling point of liquid hydrogen (20.4~K). The standard method 
of hydrogen isomer separation consists of cooling down the gas to 
20.4~K in the presence of a catalyst (like activated charcoal or 
Fe(OH)$_{3}$) which speeds up the equilibration. After equilibrium 
is reached, 99.8\% of the hydrogen molecules are in the lowest 
rotational state $J$=0 which is the para state. After warming up 
the gas, one has pure para hydrogen at ambient temperatures  because 
of extremely slow ortho-para conversion.

The discovery  of hydrogen spin isomers triggered extensive 
investigations into their physical, chemical and even biological 
properties. Contrary to the intuitive feeling that nuclear spins 
are ``deeply hidden''
inside the molecule and cannot be important in practice, they play 
a decisive role in some practical problems. The most famous example 
is the storage of liquid hydrogen, e.g., as a rocket fuel, where 
considerable boil-off is caused by the energy released from 
ortho-para conversion. Research on hydrogen spin isomers, over 
nearly 70 years, has been reviewed in a few monographs 
\cite{Farkas35,Cremer43,Trapnell55,Schmauch64,Ilisca92} 
and is not considered in this paper.

\subsection{Other molecules}

The study of molecular spin isomers gave the first experimental 
background for the concept of nuclear spin (see, e.g., 
\cite{Heitler29}), finally establishing the general relation 
between spin and statistics of identical particles. For molecular 
physics this theorem has as a consequence that molecules having 
identical nuclei in symmetrical positions (e.g., H$_2$, NH$_3$, 
CH$_4$, N$_2$, etc.) occur in nature only in the form of nuclear 
spin isomers for which selection rules prescribe
particular rotational quantum numbers to particular spin states.

Nuclear spin isomers of molecules other than hydrogen formed a 
long standing puzzle. Even 50 years after the discovery of hydrogen 
spin isomers one could safely state that ``no one has ever produced 
gaseous samples of any molecules other than H$_2$ or D$_2$ in which 
the ratio of the concentrations of the different nuclear spin 
symmetry species is different from the high temperature equilibrium 
value'' \cite{Bloom72}.  Thus, on one hand, quantum mechanics 
predicted the existence of stable spin isomers (see, for example, 
\cite{Landau81}, p.~426), and  on the other hand, almost nothing 
was known about their stability and their properties, for lack of 
practical separation methods.
The simple technique of hydrogen isomer separation obviously fails 
in the case of heavy molecules, which have much smaller rotational 
level splitting and a much higher boiling point. Consequently, at 
temperatures sufficiently low for isomer enrichment the gas becomes 
solid. Thus the separation of spin isomers of heavy molecules 
requires a special technique. This is a non-trivial problem because 
spin isomers have identical masses and quite similar physical and 
chemical properties. Nevertheless, separation methods are under 
development now. In this Section we briefly review these methods.

$\bullet$~{\underline {Low-temperature solids.}} Small molecules 
embedded in special matrices partially retain their spin isomer 
features. This allows one to produce in solids at low temperature 
significant enrichment of spin isomers in comparison with their 
abundance at room temperature. One may hope that the
enrichment will be retained after fast evaporation of these molecules 
into the gas phase. In paper \cite{Curl66JCP} an attempt was
made to separate spin isomers of methane using rapid heating of a 
solid held at low temperature. In a subsequent paper \cite{Curl67JCP} 
(see
also \cite{Ozier67JCP}), these authors attributed the negative 
result of \cite{Curl66JCP} to fast conversion due to the degeneracy 
of methane states having different spin symmetry. Yet this issue 
deserves further investigation because fast (although in a different 
time range) cooling of methane in a molecular jet is not accompanied 
by spin conversion \cite{Amrein88JMS,Hepp94JMS,Georges98JMS}.

$\bullet$~{\underline{Selective photolysis.} Due to the difference in 
allowed rotational quantum numbers, different spin
isomers are spectroscopically distinguishable. (In fact, this is the 
essence of the famous line intensity alternation effect). In the 
enrichment method proposed in \cite{Curl67JCP} narrow band laser 
radiation
destroys one of the spin isomers from an equilibrium mixture by 
photolysis. This method was first applied to the enrichment of 
I$_{2}$ spin isomers \cite{Balykin76CP}. It is not certain yet 
if I$_{2}$ can be enriched by selective photolysis because the 
results \cite {Balykin76CP} could not be reproduced in \cite{Booth89CP}.

Selective photolysis was used for enrichment of spin isomers of 
formaldehyde (CH$_{2}$O) molecules in \cite{Schramm83CPL}. The 
authors obtained enriched  hydrogen (one of the photolysis products), 
but enrichment of CH$_{2}$O was not achieved. The conclusion was 
that the spin conversion is too fast in comparison with the duration 
of the photolysis \cite{Schramm83CPL}. Later, with an improved setup, 
enrichment of CH$_{2}$O spin isomers by selective photolysis was 
nevertheless demonstrated \cite{Kern89CPL}. Those authors measured 
the life time of the formaldehyde spin isomers to be 200 sec. As a 
possible mechanism of conversion the authors \cite{Kern89CPL} proposed 
the mixing of states model \cite{Curl67JCP}.

$\bullet$~{\underline{Exchange reactions.} Another approach to the 
separation of spin isomers of heavy molecules is based on the use 
of chemical exchange reactions with spin-polarized atoms. This 
technique proved feasible for the enrichment of nuclear spin isomers 
of diatomic molecules in \cite{Weber74PL,He90JCP} where enrichment 
of Na$_2$ and Li$_{2}$ was obtained. 

$\bullet$~{\underline{Selective adsorption/condensation.} The 
difference in rotational quantum numbers for different spin 
isomers can cause a difference in their physical properties, e.g., 
in adsorption on surfaces. This can be used for separation. Thus 
spin isomers of water molecules were separated by selective 
condensation \cite{Konyukhov86} and by selective adsorption on 
Al$_{2}$O$_{3}$ surface \cite{Konyukhov88}. This group measured 
the life time of water spin isomers to be $4.4\pm0.2$~days. 
Presumably, the isomers convert on  the container surface 
\cite{Konyukhov88}. No model was proposed so far for the conversion 
of water.

$\bullet$~{\underline{Chemical reactions.} The dependence of chemical 
reactions on the nuclear spin state of the reacting molecules was 
predicted in \cite{Quack77MP}. This
selectivity was employed in experiment \cite{Oka97PRL} to obtain 
enriched samples of H$_{3}^{+}$ molecules by chemical reactions 
involving enriched H$_2$.

$\bullet$~{\underline{Light-induced drift.} A breakthrough in the 
separation of spin isomers was achieved in Ref.~\cite{Krasnoperov84JETPL} 
by using Light-Induced Drift (LID) \cite{Gel79JETPL}. CH$_3$F isomer 
separation was analogous to the previously  performed enrichment of 
the CH$_3$F isotope species  (see \cite{Panfilov83JETP} and 
references therein). The LID effect proved to be a convenient tool 
since it enables one to separate spin isomers in the gas phase at 
well-defined gas composition, temperature and pressure. Although 
this technique can be applied to any molecular species in principle, 
the first experiments were performed on CH$_3$F for the following 
reasons. The two nuclear spin species, characterized by their total 
proton spin $I$=3/2 or 1/2, have well-distinguishable rovibrational 
absorption lines which are easily accessible by a CO$_2$ laser. In 
addition, CH$_3$F is chemically inactive and has low adsorption onto 
the cell walls.
These fortunate circumstances have promoted substantial progress in 
the field. The first measurement of the life time of nuclear spin 
isomers of polyatomic molecules was performed for $^{12}$CH$_{3}$F 
and gave 2~hours  \cite{Chap85CP}. 
In subsequent measurements \cite{Bakarev86JETPL} nearly two orders of
magnitude difference in the conversion rates of spin isomers of 
$^{13}$CH$_{3}$F and $^{12}$CH$_{3}$F molecules was observed. 

At the time of these experiments the nuclear spin conversion of 
CH$_{3}$F molecules was considered to be a heterogeneous process.
The situation changed  when it was proven \cite{Chap90JETP} that 
the  CH$_3$F conversion is a gaseous process having an anomalously 
large isotope dependence  
$(14\pm 1)\times 10^{-3}$\ s$^{-1}$/torr for the $^{13}$CH$_{3}$F 
isomers and 
$(0.31\pm 0.03)\times 10^{-3}$\ s$^{-1}$/torr
for the $^{12}$CH$_{3}$F isomers. Although these rates are extremely 
fast if compared to the H$_2$ case, they are astonishingly slow if 
considered from a gas-kinetic point of view: the molecules retain 
their spin state even after suffering some 10$^9$ collisions which 
totally scramble the rotational state. The first feeling of a 
researcher in this situation is that the phenomenon is too complicated 
to expect a simple description based on first principles. Nevertheless, 
there is such simple model as will be shown below.

\section{THEORY}

\subsection{Quantum relaxation}

An explanation of the CH$_{3}$F spin conversion was suggested 
\cite{Chap90JETP} in the framework of a model based on intramolecular 
mixing of
ortho and para states of the molecule. This model
was proposed in a theoretical paper by Curl et al \cite{Curl67JCP} as  
a tentative mechanism of spin conversion in H$_{2}$O, CH$_{2}$O and 
CH$_{4}$. The model \cite{Curl67JCP} formed the basis of a new 
approach to the spin conversion in polyatomic molecules.

The essence of spin conversion induced by intramolecular mixing 
consists of the following. Let us divide the quantum states of a 
molecule into two subspaces, ortho and para, each having their own 
rotational energy levels. This is symbolized in Figure~\ref{subs}. 
Let us assume for simplicity that only one pair of ortho and para 
states ($m$-$n$) is mixed by an intramolecular perturbation. Suppose 
that at the beginning of the conversion process a test molecule is 
placed in the ortho
subspace. Collisions of the test molecule with the surrounding 
particles cause fast migration inside the ortho subspace but cannot, 
by assumption, change its spin state directly. This running up and down
along the ladder of ortho states continues until the molecule arrives 
in the state $m$ which is mixed with the energetically close para 
state $n$. During the free flight after
this collision, the para state $n$ will be admixed by the internal 
perturbation to the ortho state $m$, which creates a coherent mixture 
of the $m$ and $n$ states. The next collision can destroy this coherent 
mixture of states by transferring the molecule to other para states. 
This will localize the molecule inside the para subspace, and conversion 
has taken place.

To summarize, the mechanism of spin conversion is based on 
intramolecular mixing of ortho and para states and collisional 
destruction of the coherence between the states.
This type of relaxation deserves a special name and can be called 
``quantum relaxation'' \cite{Chap96PA}. Processes of similar origin 
are well known in various parts of physics. First of all we can 
mention the relaxation of magnetic polarization in NMR 
\cite{Bloembergen61}. Another example is the vibrational energy 
transfer enhanced by Coriolis perturbation (see \cite{Orr95CP} 
and references therein). The same mechanism is responsible for 
the singlet-triplet relaxation in spin-radical pairs \cite{Salikhov84}. 
Analogous mechanisms can be found also in elementary particle 
physics, e.g., neutral kaon decay \cite{Feynman}.

Below we give a rigorous model of isomer conversion by quantum 
relaxation \cite{Chap90JETP}, hoping that the reader will enjoy 
seeing how a few steps of simple mathematics bring us to a general 
solution of a complicated problem. Suppose, that the molecular 
Hamiltonian is the sum of the two parts 
\begin{equation}
\hat{H}=\hat{H}_{0}+\hbar \hat{V},  \label{H}
\end{equation}
where $\hat{H}_{0}$ is the main part of the Hamiltonian which has 
pure ortho and para states as the eigen states; $\hat{V}$ is a small 
intramolecular perturbation which mixes the ortho and para states.

The Liouville equation for the density matrix $\rho $ of the molecule 
in the representation of the eigen states of the operator $\hat{H}_{0}$ 
reads 
\begin{equation}
(\partial/\partial t) \rho _{\alpha \alpha_1}=S_{\alpha \alpha
_{1}}-i\,[\,\hat{V},\rho \,]_{\alpha \alpha _{1}}\ ,  \label{rho}
\end{equation}
where $S_{\alpha \alpha _{1}}$ is the collision integral and $\alpha $ 
and $\alpha _{1}$ represent the complete sets of quantum states.

Using Eq.(\ref{rho}) one can obtain an equation which governs the 
change of molecular concentration in one particular nuclear spin state. 
This can be done by calculating the trace of Eq.(\ref{rho}) over all 
states of one spin isomer. For example, for the concentration of ortho 
molecules ($\rho _{o}$) one has 
\begin{equation}
(\partial/\partial t)\rho_o=2Re\sum_{\alpha \in o,\alpha ^{\prime
}\in p}i\rho _{\alpha \alpha'}V_{\alpha'\alpha };\quad
\rho _{o}\equiv \sum_{\alpha \in o}\rho _{\alpha \alpha },  \label{rho_o}
\end{equation}
where $\alpha $ runs over all ortho ($o$) states and $\alpha^{\prime}$ 
runs over all para ($p$) states. Henceforth, unprimed quantum numbers 
will refer to the
ortho states, and the primed ones to the para states.

When deriving Eq.(\ref{rho_o}) we have assumed  that collisions
conserve the number of molecules in each spin state (no ``direct'' 
conversion). This can be expressed as 
\begin{equation}
\sum_{\alpha \in o}S_{\alpha \alpha }=\sum_{\alpha'\in
p}S_{\alpha'\alpha'}=0\ .  
\label{s}
\end{equation}
This property means that the collisional cross-section of ortho-para 
transfer is equal to zero: $\sigma(p|o)$=0.
The justification of the relations (\ref{s}) for CH$_{3}$F molecules, 
both experimental and theoretical, can be found in 
\cite{Nagels95JCP,Nagels96PRA,Chap96CPL}. 

To find the time dependence of $\rho _{o}$ from Eq.(\ref{rho_o}) we 
need to know the off-diagonal matrix elements of $\rho$,
which are a measure of the coherence between ortho and para states. 
From Eq.(\ref{rho}) we have 
\begin{equation}
(\partial/\partial t)\rho _{\alpha \alpha'}=
S_{\alpha\alpha'}-i\,[\,\hat{V},\rho \,]_{\alpha \alpha'}\ .  
\label{rho1}
\end{equation}

In first order perturbation theory, on the right hand side of 
Eq.(\ref{rho1}) one has to keep only the unperturbed values of the 
density matrix. Those are the diagonal matrix elements of $\rho $. 
This leads to the equation 
\begin{equation}
(\partial/\partial t)\rho _{\alpha \alpha'}=S_{\alpha
\alpha'}-iV_{\alpha \alpha'}(\rho _{\alpha ^{\prime
}\alpha'}-\rho _{\alpha \alpha })\ .  \label{rho2}
\end{equation}
Here we have assumed that perturbation $\hat{V}$ has no diagonal matrix
elements. Such definition of $\hat{V}$ is always possible by proper
separation (\ref{H}) of the molecular Hamiltonian.

Further, we will model the off-diagonal elements of the collision 
integral solely by a decay process: 
$S_{\alpha \alpha'}=-\Gamma _{\alpha \alpha ^{\prime
}}\rho _{\alpha \alpha'}$. The decoherence rate $\Gamma _{\alpha\alpha'}$ 
is approximately equal to the average of the population
decay rates in the states $\alpha $ and $\alpha'$. Note that 
$\Gamma_{\alpha\alpha'} $ is proportional to the gas pressure. 
Moreover, we assume all $\Gamma _{\alpha \alpha'}$ to be equal: 
$\Gamma _{\alpha \alpha'}\equiv \Gamma $. 
Thus the off-diagonal terms of the collision integral will be 
modeled by the {\sl Anzatz} 
\begin{equation}
S_{\alpha \alpha'}=-\Gamma \rho _{\alpha \alpha'}\ .  
\label{s1}
\end{equation}
The validity of this model will be tested below by comparison with 
experiment.

The intramolecular perturbation $\hat{V}$ is time independent. 
Consequently, the time dependence of $V_{\alpha \alpha'}$ is given by  
$\exp(i\omega _{\alpha \alpha'}t)$  where 
$\hbar \omega _{\alpha \alpha^{\prime }}$ 
is the energy gap between the states 
$\alpha $ and $\alpha^{\prime }$. 
Using the assumptions made above, the steady-state solution of 
Eq.(\ref{rho2}) reads 
\begin{equation}
\rho_{\alpha \alpha'}=\frac{-iV_{\alpha \alpha'}}{\Gamma +
i\omega _{\alpha \alpha'}}(\rho _{\alpha'\alpha'}-\rho_{\alpha\alpha})\ .  
\label{rho3}
\end{equation}
Combining (\ref{rho_o}) and (\ref{rho3}) one has 
\begin{equation}
\frac{\partial \rho _{o}}{\partial t}=
\sum_{\alpha \in o,\,\alpha'\in p}
\frac{2\Gamma \mid V_{\alpha \alpha'}\mid ^{2}}{\Gamma
^{2}+\omega _{\alpha \alpha'}^{2}}(\rho _{\alpha'\alpha
^{\prime }}-\rho _{\alpha \alpha })\ .  \label{drho_o}
\end{equation}

The diagonal elements of the density matrix (which are populations 
of states) are determined by the Boltzmann distribution, $W_{\alpha }$, 
for ortho and para isomers independently: 
\begin{equation}
\rho _{\alpha \alpha }=\rho _{o}W_{\alpha };\quad 
\rho _{\alpha'\alpha'}=\rho _{p}W_{\alpha'},  
\label{Boltz}
\end{equation}
because the rotational relaxation inside the ortho and para subspaces
is many orders of magnitude faster than the ortho-para conversion. 
In Eq.(\ref{Boltz}) $\rho _{o}$ and $\rho _{p}$ are the total 
concentrations of ortho and para molecules, respectively. The Boltzmann 
factors are determined in the standard way by the expression
\begin{equation}
W_{\alpha }=Z^{-1}_{ortho}\exp (-E_{\alpha}/kT) ,  
\label{W}
\end{equation}
for the ortho molecules  and similarly for para. In (\ref{W}), 
$E_{\alpha }$ is the energy of state $\alpha $; $Z_{ortho}$ and 
$Z_{para}$ are the partition functions for ortho and para molecules, 
respectively. 

Let us represent $\rho _{o}$ as the sum of a steady-state and a
time-dependent part: $\rho _{o}=\overline{\rho }_{o}+\delta \rho _{o}(t)$.
By taking into account that the total molecular concentration $N=\rho
_{o}+\rho _{p}$ is conserved, one finds from (\ref{drho_o}) and 
(\ref{Boltz}) an exponential decay: 
\begin{equation}
\delta \rho _{o}(t)=\delta \rho _{o}(0)e^{-\gamma t}\ ,  \label{drho}
\end{equation}
with the conversion rate  
\begin{equation}
\gamma =\sum_{\alpha\in o,\alpha'\in p}
\frac{2\Gamma \mid V_{\alpha\alpha'}\mid^2}
{\Gamma ^{2}+\omega _{\alpha\alpha'}^2}
\left(W_{\alpha'}+W_{\alpha}\right).
\label{gamma}
\end{equation}
This expression gives the solution to the problem in first order 
perturbation theory. It is valid for not too strong mixing, such that 
\begin{equation}
     \mid V\mid^2 \ll \frac{\nu_{rot}}{4\Gamma}\max\{\Gamma^2,\omega^2\},
\label{1st}
\end{equation}
where $\nu_{rot}$ is the rotational relaxation rate. Conversion in 
the more general case is considered in \cite{Chap96PA}.

\subsection{Ortho-para state mixing in CH$_{3}$F}

The model of spin conversion described above is simple and clear, 
but is there something inside a symmetrical molecule which mixes 
the ortho and para states?  We know now that such mixing can be 
produced by hyperfine interactions. These interactions are very 
weak and usually show up as small effects on a much larger 
background. Quantum relaxation of spin isomers provides an 
example of a phenomenon for which hyperfine interactions are 
the leading force.

Hyperfine interactions in molecules are important for various 
physical problems, first of all for hyperfine spectroscopy. We 
note that for the nuclear spin conversion one needs a specific 
part of hyperfine interaction which  mixes the ortho and para states. 
This part of hyperfine interaction is unimportant for standard 
hyperfine spectroscopy but is important for level 
crossing/anticrossing spectroscopy (see Ref.~\cite{Ozier81CJP} 
and references therein).

Let us recall the classification of CH$_3$F quantum states. 
As a consequence of the CH$_3$F symmetry the molecules exist in 
form of two nuclear spin isomers: ortho and para which have a 
total spin of three hydrogen nuclei $I=3/2$ and $I=1/2$, 
respectively (see, for example \cite {Townes}). The ortho 
molecules have $K$-values ($K$ refers to the angular momentum 
projection on the molecular symmetry axis) divisible by 3: 
$K=0,3,6\ldots $. For para isomers only $K=1,2,4,5\ldots $ are 
allowed. Consequently, the quantum states of CH$_{3}$F are 
divided into two subspaces, ortho and para,  which are shown 
in Figure~\ref{subs} for the particular case of $^{13}$CH$_3$F 
molecules. 

An analysis of ortho-para mixing in CH$_{3}$F in relation with 
the spin conversion problem was performed in a few papers. Two 
sources of mixing in CH$_3$F have been considered so far: 
spin-spin interactions between the molecular nuclei 
\cite{Chap91PRA,Guskov95JETP,Bahloul98PhD} and spin-rotation 
interactions 
\cite{Guskov95JETP,Bahloul98PhD,Chap96,Bahloul98JPB,Ilisca98PRA}. 

{\underline{Spin-spin interactions.}
The ortho and para states in CH$_3$F are mixed by spin-spin 
interaction between the three hydrogen nuclei ($\hat{ V}_{HH}$), 
the fluorine--hydrogen nuclei ($\hat{V}_{FH}$) and the 
carbon--hydrogen nuclei ($\hat{V}_{CH}$) in the case of 
$^{13}$CH$_3$F (see Figure~\ref{coord}). These interactions 
can be written as 
\begin{eqnarray}
\hat{V}_{HH} &=&P_{HH}\sum_{m<n}{\bf \hat{I}}^{(m)}{\bf \hat{I}}
^{(n)}\ ^\bullet_\bullet\ {\bf T}^{(m,n)};\quad 
\hat{V}_{FH}=P_{FH}\sum_{m}{\bf \hat{I}}
^{(m)}{\bf \hat{I}}^{F}\ ^\bullet_\bullet\ {\bf T}^{mF}\ ;\nonumber\\
\quad \hat{V}_{CH} &=&P_{CH}\sum_{m}{\bf \hat{I}}^{(m)}{\bf \hat{I}}
^{C}\ ^\bullet_\bullet\ {\bf T}^{mC}\ ,  
\label{ss}
\end{eqnarray}
with $m,n=1,2,3$ denoting the protons.
In (\ref{ss}) $P$ are scale factors having the order of magnitude 
$10^{4}$ Hz. Their numerical values are given in \cite{Chap91PRA}; 
$\hat{{\bf I}}$ are the spin operators of the corresponding nuclei; 
${\bf T}$ are the second rank tensors for the magnetic dipole--dipole 
interaction having the form, e.g.,
\begin{equation}
T_{ij}^{mF}=\delta _{ij}-3n_{i}^{mF}n_{j}^{mF},  
\label{TmF}
\end{equation}
where ${\bf n}^{mF}$ is the unit vector directed from the H$^{(m)}$ 
to the F nucleus. 

The spin conversion rate in $^{13}$CH$_{3}$F molecules induced by the
spin--spin interactions $\hat{V}^{SS}=\hat{V}_{HH}+\hat{V}_{FH}+
\hat{V}_{CH}$,
is given by the expression (\ref{gamma}) having a mixing efficiency 
for the ortho-para level pair ($J'$,$K'$)-($J$,$K$) \cite{Chap91PRA} 
\begin{eqnarray}
\sum\mid V^{SS}_{\alpha\alpha'}\mid^2 &\equiv& F_{SS}(J',K'|J,K) 
=\,(2J^{\prime }+1)(2J+1)\left( 
\begin{array}{rcr}
 J' & 2 & J \\ 
-K' & q & K
\end{array}
\right) ^{2}\times   \nonumber \\
&&\left( 3\mid P_{HH}{\cal T}_{2q}^{(1,2)}\mid ^{2}+2\mid P_{FH}{\cal T}
_{2q}^{1F}\mid ^{2}+2\mid P_{CH}{\cal T}_{2q}^{1C}\mid ^{2}\right) .
\label{FSS}
\end{eqnarray}
Here ${\cal T}$ are the spherical components of the corresponding 
${\bf T}$-tensors calculated in the molecular frame (for their 
numerical values see \cite{Chap91PRA}); (:\ :\ :) stands for
the 3-j symbol. In the left-hand side of (\ref{FSS}) summation is 
performed over 
degenerate quantum numbers of the states $\alpha$ and $\alpha'$ which 
are the projections on the laboratory quantization axis of molecular 
angular momentum
$M$, spins $\sigma$, $\sigma^F$, $\sigma^C$ of the three protons, 
fluorine and carbon nucleus, respectively. 

The selection rule for the spin-spin mixing follows from 
Eq.(\ref{FSS}):
\begin{equation}
\mid \Delta J\mid \ \leq 2\ ;\ \ \ J'+J\geq2\ ;\ \ \ \ 
\mid \Delta K\mid \ \leq 2\ .
\label{rulesss}
\end{equation}
Note that mixing of states having $\mid \Delta K\mid =0$ does not
contribute to the spin conversion. By contrast, the level shift 
in hyperfine spectroscopy is dominated by the diagonal matrix 
elements of $\hat{V}$ having $\mid \Delta K\mid =0$.

The spin-spin interaction in CH$_3$F can be determined with 
rather high accuracy, limited presently to a few percent by the 
uncertainty in the molecular spatial structure. The most accurate 
data for the CH$_3$F structure are given in Ref.~\cite{Egawa87JMSt}.

{\underline{Spin-rotation interaction.} Another source of 
ortho-para mixing in CH$_{3}$F is the spin-rotation coupling 
which results from the interaction of the nuclear
spins with the magnetic field induced by molecular rotation 
\cite{Guskov95JETP,Bahloul98PhD,Chap96,Bahloul98JPB,Ilisca98PRA}. 
The spin-rotation perturbation relevant for the ortho-para mixing 
can be written as \cite{Ilisca98PRA}
\begin{equation}
\hat V_{SR} = \frac{1}{2}\sum_n\left[\hat{\bf J}\bullet{\bf C}^{(n)}
               \bullet\hat{\bf I}^{(n)} + H.C.\right],
\label{SR}
\end{equation}
where ${\bf C}^{(n)}$ is the spin-rotation tensor for the $n$-th 
hydrogen nucleus. The expression for the ortho-para mixing 
efficiency due to the spin-rotation interaction can be found 
in the original papers 
\cite{Guskov95JETP,Bahloul98PhD,Chap96,Bahloul98JPB,Ilisca98PRA}.

The selection rules for the spin-rotation interaction read 
\begin{equation}
\mid \Delta J\mid \ \leq 1\ ;\ \ \ \mid \Delta K\mid \ \leq 2\ .
\label{rulessr}
\end{equation}
As can be seen from the selection rules (\ref{rulesss}) and 
(\ref{rulessr}), there are ortho-para level pairs which are 
mixed exclusively by spin-spin but not by spin-rotation interaction, 
viz., $\mid\Delta J\mid=2$. This can be used to disentangle the 
contribution to the spin conversion from these two mechanisms of 
state mixing \cite{Bahloul98JPB}.

In contrast to spin-spin interaction, the spin-rotation interaction 
in CH$_{3}$F is known presently only approximately. It was proposed 
\cite{Guskov95JETP,Bahloul98JPB} to use nuclear spin conversion itself 
as a source of information on spin-rotation perturbation in molecules. 
Such information would be complementary to the information provided by 
standard hyperfine spectroscopy.

\subsection{CH$_{3}$F level structure and theoretical conversion rates}

Spin conversion by quantum relaxation is dependent on the 
position of ortho and   para states, notably the
ortho-para level gaps. They are not accessible directly by standard
spectroscopical methods. Nevertheless, if a complete set of molecular
parameters is available one can calculate the positions of all 
states and thus the ortho-para level gaps. The accuracy of the 
molecular parameters should be rather high to guarantee the 
determination of the level position within $ 1-10 $ MHz because 
only close ortho-para level pairs contribute substantially to 
the conversion. For CH$_{3}$F such accurate parameters are 
available from high-resolution spectroscopy 
\cite{Graner76MP,Lee87JMS,Papousek93JMS33,Papousek93JMS62,Papousek94JMS}. 

A search for close ortho-para level pairs in CH$_{3}$F was 
performed in \cite{Chap91PRA}. At the time, a complete set 
of ground state molecular parameters was available for 
$^{12}$CH$_{3}$F only \cite{Graner76MP}. For $^{13}$CH$_{3}$F 
the molecular parameters $A_{0}$ and $D_{0}^{K}$ were missing. 
Nevertheless, a search for close ortho-para level pairs was 
performed  for both molecules using the estimation of 
$A_0$=5.18240(6)~cm$^{-1}$ made by T~Egawa and K~Kuchitsu 
(unpublished data) and assuming $D_{0}^{K}$ equal for both isotopes. 
This search revealed that only two ortho-para level pairs are 
important for the spin conversion in $^{13}$CH$_{3}$F. For  
$^{12}$CH$_{3}$F four important ortho-para level pairs were found. 
This choice of level pairs was confirmed and their energy gaps 
determined more precisely in 
\cite{Chap93CPL} on the basis of new accurate molecular parameters 
determined in \cite{Papousek93JMS33,Papousek93JMS62}. 
The best presently
available values for the CH$_3$F ortho-para level gaps are given 
in Table~1.

In addition to the ortho-para level spacing, $\omega_{\alpha\alpha'}$, 
and the efficiency of the ortho-para mixing, $V_{\alpha\alpha'}$, 
the decoherence rate, $\Gamma $, should be determined in order to 
calculate the conversion rates. This is a rather complicated 
problem that has not yet been resolved. Approximately, the rate 
$\Gamma$ can be taken equal to the level population decay rate. 
Even so, the uncertainty remains because the population decay 
rates of the states important for the conversion (see Table~1) 
have not been measured. There is a measurement of the population 
decay rate in the state ($J$=4, $K$=3) of $^{13}$CH$_{3}$F 
\cite{Jetter73JCP} which was determined to be 
1$\cdot $10$^{8}$~s$^{-1}$/torr.

Since the decoherence rate $\Gamma $ is close to but nevertheless 
different from the level population decay rate, it is more 
consistent to determine $\Gamma$ from the nuclear spin conversion 
itself by fitting the measured conversion rate. Such an approach 
\cite{Nagels96PRL,Nagels98PRA} yields
$\Gamma \simeq1.8\cdot $10$^{8}$~s$^{-1}$/torr which is indeed 
close to the level population decay rate 1$\cdot$10$^{8}$~s$^{-1}$/torr 
\cite{Jetter73JCP}.
The various mixing processes in CH$_{3}$F are illustrated in Table~1
where the contributions to the conversion are calculated assuming 
the decoherence decay rate $\Gamma=1.8\cdot $10$^{8}$~s$^{-1}$/torr 
for all pairs of ortho-para states in both molecules.

From the data in Table~1 one can see that the conversion in 
$^{12}$CH$_{3}$F by spin-spin interaction is much slower than 
in $^{13}$CH$_{3}$F. On the other hand, in $^{12}$CH$_{3}$F the 
spin-rotation interaction was predicted to be the leading mechanism 
\cite{Guskov96,Ilisca98PRA}. The key point is the right choice of 
the decoherence rate $\Gamma$ for the most important level pair 
in $^{12}$CH$_{3}$F:
(28,5)-(27,6). It was proposed in \cite{Ilisca98PRA} to use  
$\Gamma\simeq2\cdot10^7$~s$^{-1}$/torr for this pair in order 
to fit the experimental value for the rate. 

\section{EXPERIMENTAL METHOD}

\subsection{Spin isomer separation by LID}

It is the introduction of the Light-Induced Drift (LID) 
\cite{Gel79JETPL} as a separation tool that has brought 
substantial progress in the field of nuclear spin isomers over 
the last decade. The essence of LID is explained in Figure~\ref{lid}. 
Let us consider a gas of two-level particles interacting with a 
traveling monochromatic wave. Due to the Doppler effect, the 
radiation will excite particles selectively with respect to their 
velocity component along the radiation  {\bf k}-vector. 

Suppose that the absorbing particles are diluted in a buffer gas 
and that the laser frequency is chosen such that the radiation 
excites only  molecules moving away from the laser. Because the 
excited molecules have in general a different (usually bigger) 
kinetic cross section than the unexcited particles, the mean free 
path for molecules moving away from the laser will be different 
(usually smaller) than for molecules moving towards the laser. 
This difference in mean free path will create a drift of absorbing 
particles towards the laser. The buffer gas will flow in the 
opposite direction as required by momentum conservation. The 
direction of the fluxes depends on the sign of the laser frequency 
detuning from the absorption line center and the direction of the 
wave vector {\bf k}. In a closed tube the LID effect results in 
a spatial separation of the two gas components, which can be quite 
substantial even if the fluxes themselves are relatively small.

A quantitative description of the LID effect in molecules is a  
complicated problem because LID depends on small difference in 
transport properties between excited and ground state molecules, 
which is rather difficult to calculate. A crucial parameter is 
the relative change in cross section, or -- more precisely -- in 
collision rate, $\Delta\nu/\nu$, upon excitation. In general, 
$\Delta\nu/\nu$ depends on velocity, and thus on the detuning. 
In some cases $\Delta\nu/\nu$ can even change sign as a function 
of detuning, giving rise to so called ``anomalous LID'' 
(see, e.g., Refs.~\cite{Meer92PRA}). Here we will limit ourselves 
to the simplified case that $\Delta\nu/\nu$ can be considered constant. 
In this case the description of LID is relatively simple 
\cite{Mironenko81IANS} and yields for the difference in absorbing 
particle density between the ends of the tube
\begin{equation}
\Delta n=-\frac{\Delta \nu }{\nu }\frac{2\Delta S}
{\hbar \omega v_{0}}\varphi(\Omega ),  
\label{dn}
\end{equation}
where $\Delta S$ is the absorbed laser intensity; $\hbar\omega$ is 
the photon energy; $v_{0}=\sqrt{2kT/m}$ is the thermal speed; 
$\varphi (\Omega )$ describes the behavior as a function of detuning, 
$\Omega =\omega -\omega _{0}$, of the laser frequency, $\omega $, 
from the absorption line center frequency, $\omega _{0}$. The spectral 
function $\varphi (\Omega )$ can be readily calculated if the 
homogeneous line width of the absorbing transition is known. For 
small frequency detuning and low pressure, the function 
$\varphi(\Omega)\simeq v_L/v_0$, where $v_L=\Omega/k$ is the 
resonant velocity component of the absorbing particle along the 
${\bf k}$-vector. Examples of $\varphi (\Omega )$ for various 
homogeneous line widths, as well as corrections to (\ref{dn}) 
due to the finite gas dilution are given in \cite{Meer89PRA}.

It was shown in Ref.~\cite{Chap89IANS} that the LID effect in 
the CH$_3$F isotope mixture, which should behave similar to the 
mixture of spin isomers, is consistent with the model of constant 
$\Delta\nu/\nu$. The change in collision rate, $\Delta \nu /\nu $, 
upon rovibrational excitation was found to be $\simeq$1\%. For more 
details on the LID effect see Ref.~\cite{Rautian91} and 
Refs.\cite{Chap89IANS,Hermans92IRPC} which review the LID effect 
in molecular gases.

A prerequisite of the LID separation of gas components is their 
spectral distinguishability. This is perfectly satisfied for CH$_3$F 
 for which ortho and para isomers are different by $K$ and thus have 
 different absorption spectra. 
There are two convenient coincidences between absorption lines of 
$^{13}$CH$_{3}$F and $^{12}$CH$_{3}$F molecules in the $\nu_3$ 
fundamental band (C--F stretch) and CO$_{2}$-laser lines 
\cite{Freund74JMS}, which were used for spin isomer separation. 
The absorption spectra of ortho and para $^{13}$CH$_{3}$F near 
the P(32) line of the 9.6$~\mu $ band of a CO$_{2}$-laser are shown 
in Figure~\ref{spectr}. As can be seen from these spectra, the 
CO$_{2}$-laser frequency tuned to the center of the 9P(32) laser 
line excites ortho $^{13}$CH$_{3}$F molecules in the blue wing of 
the R(4,3). This results in LID of ortho isomers towards the laser. 
In the case of $^{12}$CH$_{3}$F, excitation of the molecular 
absorption line $Q(12,2)$ by the 9P(20) laser line induces a 
drift of para molecules away from the laser. In a typical 
experiment on CH$_3$F spin isomer an enrichment of 10\% was 
produced, sufficient to have good signal-to-noise ratio.

\subsection{Relaxation time measurement}

The experiments on CH$_3$F  were performed in Novosibirsk 
and Leiden using similar approaches. Separation 
of spin isomers was achieved in a long and thin glass tube using 
LID as outlined in Section~3.1.  One end of
the separation tube was connected with a test cell to collect an 
enriched gas sample. The other end of the separation tube was 
connected with a ballast
volume, thus providing an equilibrium reference sample 
(see Figures~\ref{novosib} and \ref{leiden}). 
The degree of enrichment (or depletion, depending on the particular 
arrangement) was monitored by absorption, using a probe laser beam 
resonant with either the ortho, or the para isomer. To increase the 
sensitivity and stability of the detection system the absorption in 
the test cell was compared differentially with the absorption in a 
reference cell having equilibrium composition.

A schematic of the setup used in Ref.~\cite{Chap90JETP} is presented 
in Figure~\ref{novosib}. The isomer separation was performed in a 
tube having  length 1.5~m and inner diameter 1.3~mm by CO$_2$-laser 
radiation (power $\simeq10~W$). The radiation frequency was 
stabilized to the CO$_2$ line center. For the enrichment detection, 
a small portion of the laser beam was directed through the test and 
reference cells which were placed on the same optical axis.
The absorption in these cells was modulated in antiphase by external 
Stark electrodes. This modulation produced a component in the probe 
beam proportional to the absorption difference between the two cells. 
This differential signal was normalized by the signal proportional 
to probe beam intensity. The latter signal was   created by additional 
Stark electrodes which modulated the absorption in the reference cell 
alone. The detection method used in \cite{Chap90JETP} has a rather 
high sensitivity but has the drawback that it can be applied to 
molecules having permanent electric dipole moment only.  

The setup used in Ref.~\cite{Nagels95JCP} is presented in 
Figure~\ref{leiden}. It has a separation tube of 30~cm length and 3~mm 
inner diameter for the study of $^{13}$CH$_3$F. For $^{12}$CH$_{3}$F 
a tube of 1~m length and 1.1~mm diameter was used to make up for the
 much smaller absorption of the $Q(12,2)$ transition used. In both 
 cases the separation laser (a CO$_2$ laser from Edinburgh Instrument, 
 Model PL5) was frequency stabilized to the CO$_2$-line center. 

For the detection, a weak probe beam from an additional (waveguide) 
laser was used. Intensity and direction of the beam were stabilized 
using an Acousto Optic Modulator with a feed back  loop and a set of 
diaphragms. The frequency was locked to the CH$_3$F absorption line 
center using an additional absorption cell with CH$_3$F gas at low 
pressure.

 It is essential to be sure that the powerful CO$_2$ laser radiation 
 indeed produces separation of CH$_3$F spin isomers and that spurious 
 effects like gas heating, laser induced thermal diffusion, etc., do 
 not contribute substantially to the signal. This important issue was 
 addressed already in the first experiment, where the spin isomer 
 separation was confirmed by comparing the optical signal with 
 additional monitoring of the gas composition by a mass-spectrometer. 
 A more decisive test was performed in \cite{Nagels98PhD} where it 
 was demonstrated, by probing various ortho and para absorption lines, 
 that the LID enrichment of ortho is accompanied by a corresponding 
 depletion of para species. We stress that the conversion of spin 
 isomers takes place in a part of the setup which is unaffected by 
 the strong laser beam used for the isomer separation.

Depending on the problem at hand, the test cell could be connected 
to a Stark cell, or to a cell at elevated temperature, or to a cell 
containing different test surfaces. This allowed to study the influence 
of various physical factors on nuclear spin conversion. 

The measurement procedure was generally as follows. First, the setup 
was filled with equilibrium gas and the detection system was set to 
zero absorption difference. Next, the enrichment was started by 
switching on the separation laser (first part of the curve in 
Figure~\ref{decay}). After a sufficient enrichment was achieved, 
the two probe cells were isolated by closing the appropriate valves 
and the conversion process began (the decay part in Figure~\ref{decay}). 
The measured decay curve was fitted by a function 
$\exp(-\gamma t) +\kappa t$, where $\gamma$ is the conversion rate, 
while the second term accounts for a (small) drift of the detection signal.

\section{EXPERIMENTAL RESULTS}

\subsection{Dependence on pressure and collision partner}

In the following Sections we review the experiments which test the 
mechanism responsible for spin conversion in CH$_{3}$F.
Let us first consider the pressure dependence. The data for  
$^{13}$CH$_{3}$F conversion in two buffer gases, 
$^{12}$CH$_{3}$F and $^{13}$CH$_{3}$F itself are shown in 
Figure~\ref{pressure}. The rates have linear pressure dependence and 
are seen to be rather close. As the reference value we give the 
conversion in pure $^{13}$CH$_{3}$F \cite{Nagels96PRA}
\begin{equation}
          \gamma_{13}/P = 
     (12.2\pm 0.6)\cdot 10^{-3}~s^{-1}/torr,
\label{leidenrate}
\end{equation}
where the subscript 13 refers to the isotopic species.

The linear pressure dependence of $\gamma_{13}$ is consistent with 
conversion by quantum relaxation. Indeed, at $P\simeq$1~torr the 
decoherence rate $\Gamma/2\pi\simeq30$~MHz, which is much smaller 
than the frequency gaps between the mixed states in $^{13}$CH$_{3}$F 
(see Table~1). In this pressure limit, where $\Gamma\ll\omega$, one 
can neglect $\Gamma^2 $ in the denominator of Eq.(\ref{gamma}) and 
the conversion rate becomes proportional to $\Gamma$, thus  proportional 
to pressure. 

Note that the linear pressure dependence of $\gamma_{13}$ at low 
pressures in itself does not allow to distinguish between conversion 
by quantum relaxation and by ordinary gaseous relaxation. The latter 
would produce conversion with 
a rate $2n\sigma (p|o)v$ which is also linear in pressure 
(see Section~5.2). The important point is that quantum relaxation is 
able to
reproduce the right order of magnitude for the conversion rate in 
$^{13}$CH$_{3}$F \cite{Chap91PRA}. If one takes as an estimation for 
the decoherence rate $\Gamma =1\cdot $10$^{8}$~s$^{-1}$/torr, as 
follows from measurement \cite{Jetter73JCP}, the spin-spin mixing 
of states in $^{13}$CH$_{3}$F gives the conversion rate 
$\gamma_{13}/P\simeq7\cdot10^{-3}$~s$^{-1}$/torr which is half the 
measured value.

Measurements of the $^{13}$CH$_{3}$F spin conversion in various buffer 
gases are presented in Figure~\ref{buff} \cite{Nagels95JCP,Nagels96PRA}. 
The conversion rates are rather close in buffer gas CH$_{3}$Cl and in 
pure CH$_{3}$F but are much smaller in buffer gases N$_{2}$ and O$_{2}$. 
As was concluded in \cite{Nagels95JCP,Nagels96PRA}, the buffer gas 
dependence of the conversion rate is in qualitative agreement with 
the buffer gas variation of the decoherence rate, $\Gamma$, which 
was estimated
in \cite{Nagels95JCP,Nagels96PRA} on the basis of pressure broadening 
data. An important result of the data in Figure~\ref{buff} is that 
the large magnetic moment of O$_2$ ($\simeq2~\mu_B$) seems to be 
unimportant the $^{13}$CH$_{3}$F spin conversion. For more details 
on this point see Section~5.

\subsection{Isotope effect}

The measurements show a large isotope effect in CH$_3$F conversion 
\cite{Chap90JETP,Nagels98PRA}. In Figure~\ref{isotope} the 
$^{12}$CH$_{3}$F conversion rate as a function of pressure is 
presented and seen to be much smaller than for $^{13}$CH$_{3}$F 
(Figure~\ref{pressure}). The ratio of the conversion rates in 
$^{13}$CH$_{3}$F and $^{12}$CH$_{3}$F measured in Leiden 
\cite{Nagels98PRA} was found to be
\begin{equation}
\gamma _{13}/\gamma _{12}=55\pm 4, 
\label{gg}
\end{equation}
which is close to the value 46$\pm$5 measured in \cite{Chap90JETP}. 
The slow conversion rate in $^{12}$CH$_3$F makes the measurements 
rather difficult to perform because of severe restrictions to the 
long term stability of the detection system.

The much smaller rate in $^{12}$CH$_{3}$F results from the bigger 
gaps between the ortho and para states in this molecule in comparison 
with $^{13}$CH$_3$F. There is one close pair of ortho-para states, viz.,
(51,4)-(50,6), but these states are situated at high energies and 
therefore hardly populated. The $^{13}$CH$_3$F conversion is 
dominated by spin-spin mixing of the ortho and para states.
In  $^{12}$CH$_{3}$F the spin-spin interaction gives only a small 
part of the observed rate \cite{Chap93CPL}. As was discussed in 
Section~2.3 and concluded in the papers \cite{Guskov96,Ilisca98PRA} 
the conversion in $^{12}$CH$_{3}$F is dominated by spin-rotation 
interaction. 

\subsection{Temperature dependence}

The levels important for spin conversion in $^{13}$CH$_{3}$F and 
$^{12}$CH$_{3}$F are situated at widely different energies. 
Consequently, one would expect quite different temperature 
dependences of the rates in the two cases as the level populations 
will be affected differently by temperature. For example, the 
population of the states (11,1) and (9,3), which are dominant 
for $^{13}$CH$_{3}$F conversion, will decrease at elevated temperature, 
in contrast with, e.g., states (51,4) and (50,6) of $^{12}$CH$_{3}$F 
whose population will increase. 

The temperature dependence of the conversion rate in the two 
molecules was reported in Ref.~\cite{Nagels98PRA}. The results of 
the measurements are presented in Figure~\ref{temp}. Indeed, the 
conversion rate in $^{12}$CH$_{3}$F grows rapidly with increasing 
temperature. By contrast, the rate in 
$^{13}$CH$_{3}$F decreases in the temperature range 300 - 600~K.

Presently we can propose only a rough model of the temperature 
dependence. The conversion rates is affected primarily through the 
Boltzmann occupation of rotational states. An additional temperature 
dependence may appear due to the temperature dependence of the 
decoherence rate $\Gamma$. It proved sufficient to apply only a 
25\% temperature variation in $\Gamma$ by the expression:
\begin{equation}
\Gamma(T)= \left(1.8-(T-T_{room})\cdot1.5\cdot10^{-3}K^{-1}\right)
\cdot10^8~s^{-1}/torr, 
\label{gT}
\end{equation}
to fit the experimental points for the $^{13}$CH$_{3}$F conversion 
in the range from room temperature $T_{room}$=297~K up to 600~K, as 
shown by the thick line in Figure~\ref{temp}.  Thus these data are 
qualitatively consistent with spin conversion by quantum relaxation 
(for $T\geq600$~K see below). 

The same temperature dependence of the decoherence rate $\Gamma(T)$ 
was used to model the conversion in $^{12}$CH$_{3}$F. Again, only 
spin-spin mixing of states was taken into account. The calculated 
rate in $^{12}$CH$_{3}$F (thin line) reproduces the overall behavior 
but fails to reproduce the magnitude of the rate. This may not be 
surprising, since an additional contribution to the conversion rate 
in $^{12}$CH$_{3}$F should result from spin-rotation interaction, as 
was pointed out in \cite{Guskov96,Ilisca98PRA}. 

Above 600~K the conversion rate in $^{13}$CH$_{3}$F increases again, 
which suggests an additional conversion mechanism at high temperatures. 
This mechanism seems to have no isotope selectivity, as one may 
conclude from Figure~\ref{temp}.

\subsection{Level-crossing resonances}

Eq.(\ref{gamma}) predicts a strong dependence of the conversion rate on
ortho-para level spacing at low pressures where $\Gamma \ll \omega $. 
To detect this effect, it was proposed \cite{Nagels95CPL} to split 
and cross the ortho and para levels of $^{13}$CH$_{3}$F by a 
homogeneous electric field. This experiment was performed in 
\cite{Nagels96PRL} with the setup in Figure~\ref{leiden} equipped 
with an additional Stark cell. 

Results of the measurement of the conversion rate in $^{13}$CH$_{3}$F 
as a function of electric field are shown in Figure~\ref{stexp}. 
As seen from these data the conversion rate is hardly affected by 
the field below 500~V/cm, but rises sharply above 600~V/cm. 

A theoretical description of the level-crossing resonances 
presented in Figure~\ref{stexp} was developed in 
\cite{Nagels95CPL,Nagels96PRL}. A homogeneous electric field 
splits each state of CH$_{3}$F into $2J+1$ magnetic sublevels 
due to the first order Stark effect. The energy gaps between the 
$M$-sublevels of the two states 
$\mid J',K',M'>$ and $\mid J,K,M>$ are given by the formula 
\cite{Townes} 
\begin{equation}
\hbar \omega _{M'M}({\cal E})=
\hbar \omega _{0}-{\cal E}d\left(\frac{M'K'}
{J'(J'+1)}-\frac{MK}{J(J+1)}\right) ,\   
\label{om}
\end{equation}
where $\hbar \omega _{0}$ is the energy gap between the states 
$\mid J',K'>$ and $\mid J,K>$ at zero electric field; ${\cal E}$
is the electric field strength, $d$ is the molecular electric 
dipole moment. For $^{13}$CH$_{3}$F, $d$ was taken to be 1.858~D 
(or $6.198\times 10^{-30}$~C$\cdot $m) \cite{Freund74JMS}.

In the model of level-crossing spectra \cite{Nagels95CPL,Nagels96PRL} 
only spin-spin interaction was taken into account. In this case each 
ortho-para level pair gives the following ``spectrum'' of $\gamma$ 
as a function of field strength:
\begin{equation}
\gamma({\cal E})=\sum_{M\in o,M'\in p}\frac{2\Gamma F_{SS}(J',K'|J,K)}
               {\Gamma ^{2}+\omega _{M'M}^{2}({\cal E})}
\left(\begin{array}{rcc}
 J' & J & 2 \\ 
-M' & M & M'-M \end{array}\right)^{2}
\left( W_{\alpha'}+W_{\alpha }\right) ,  
\label{gE}
\end{equation}
where $F_{SS}(J',K'|J,K)$ is the mixing efficiency from (\ref{FSS}). 
The selection rule 
$\mid \Delta M\mid \leq 2$ follows directly from (\ref{gE}).
The only unknown parameter in expression (\ref{gE}) is the decoherence 
rate $\Gamma$. This parameter was determined in \cite{Nagels96PRL} by 
fitting the experimental value of $\gamma$ in $^{13}$CH$_{3}$F at 
zero electric field which gave $\Gamma\simeq1.8\cdot 10^{8}$~s$^{-1}$/torr.

Once $\Gamma $ is determined, there are no free parameters in the 
model and the level-crossing spectra can be calculated. The result 
is presented in Figure~\ref{stexp}. As can be seen from these data, 
the model reproduces the main features of the measured spectrum very 
well: positions of the peaks within a few MHz as well as their 
amplitude within 10\%.

The spectrum in Figure~\ref{stexp} results from crossing of sublevels 
of the ortho state (9,3) and the para state (11,1). The level pair 
(21,1)-(20,3) gives  a constant contribution to $\gamma({\cal E})$ 
in the electric field range of Figure~\ref{stexp} because of the 
much larger gap for this level pair: 350~MHz. The states (11,1) 
and (9,3) have $\Delta J=2$ and consequently are mixed by spin-spin 
interactions but not by spin-rotation interaction. This justifies 
taking only spin-spin interaction into account in the modeling of 
the level-crossing resonances \cite{Nagels95CPL,Nagels96PRL}. 

\subsection{Quantum Zeno effect induced by collisions}

In the high pressure limit, where $\Gamma \gg \omega $, one can 
neglect $\omega $ in the denominator of (\ref{gamma}) and the 
conversion rate is seen to be inversely proportional to pressure 
(``$1/P$ dependence''). Consequently, at high pressures one has a 
very interesting regime of spin conversion in which rapid collisions 
prevent the molecule from changing its spin state: collisional 
inhibition of spin conversion. This slowing down of the conversion 
should start in the case of $^{13}$CH$_{3}$F at pressures well 
above 10~torr. Unfortunately it was not possible to perform 
measurements at such high pressures because of a few experimental 
problems. First, the LID effect is reduced at high pressures due 
to loss of velocity selectivity by pressure broadening. Second, 
the attained level of enrichment at high pressures becomes smaller 
also because the diffusion-limited separation time increases whereas 
the spin relaxation time is rather short at 10~torr, making 
back-conversion during the enrichment stage more severe. Third, 
the sensitivity of the detection system decreases because of 
overlapping ortho and para absorption lines. Another approach to 
observe the slowing down of the conversion would be to enrich at 
low pressures and to increase the pressure afterwards. Such an 
attempt was unsuccessful mainly because of spurious effects accompanying 
the compression \cite{Nagels98PhD}.

To overcome these problems one could, as an alternative, narrow the 
gap between the ortho and para levels by an external electric field. 
This would shift the slowing down of the conversion rate to lower 
pressures.
The idea was realized in  \cite{Nagels97PRL} by choosing the field 
strength at which the strongest peak in the spectrum of 
Figure~\ref{stexp} occurs. This peak is due to the crossing of 
sublevels $M'$=11 and $M$=9 at electric field 652.8~V/cm.
If this resonant condition is chosen and the spin conversion is 
dominated by  mixing of the resonant level pair, the description 
of the conversion is given by just one term from Eq.(\ref{gE}) 
which has $\omega=0$ in the denominator
and gives a 1/P dependence.

The result of the measurements is shown in Figure~\ref{zeno}. 
Indeed, there is a rapid slowing down of the conversion rate when 
the pressure increases from 0.05~torr to 0.5~torr. The plateau 
above 0.5~torr is the result of the neighboring $M'-M$ level pairs 
which do not have zero gaps at the chosen electric field and, 
consequently, give a contribution which is still increasing with 
pressure. At still higher pressures, the conversion rate would go 
down again.  

The experimental results in the high pressure range are in good 
agreement with the theory (solid line in Figure~{\ref{zeno}). The 
calculation of the pressure dependence was performed on the basis 
of Eq.(\ref{gE}) by taking into account all pairs 
($J'$=11, $K'$=1, $M'$)-($J$=9, $K$=3, $M$); $\mid M'-M\mid\le2$. 
The decoherence rate was taken to be $\Gamma$=1.75x10$^8$~s$^{-1}$/torr 
for all level pairs. Only the spin-spin mixing of states was taken 
into account because this pair of states has $|\Delta J|$=2 and, 
consequently, is not mixed by the spin-rotation interaction, see 
(\ref{rulessr}).

At low pressure the theoretical conversion rates are systematically 
larger than the measured one. This can be attributed to the saturation 
effect which should occur at low pressure in case of  degenerate 
ortho and para states when the first order perturbation theory 
breaks down \cite{Chap96PA,Chap97}

The inhibition of the conversion rate at increasing collision rate 
can be considered as an example of the Quantum Zeno effect 
\cite{Misra77JMP}. In the original version of this effect 
\cite{Misra77JMP} the inhibition of the quantum system decay is 
due to frequent measurements which destroy quantum coherence by 
projecting the system to a particular quantum state. In our case 
the decoherence is produced by molecular collisions which destroy 
the coherence between the ortho and para states in the molecule.

\section{DIRECT ORTHO-PARA TRANSITIONS IN CH$_3$F}

The investigation of nuclear spin conversion in CH$_3$F would not 
be complete without studying direct ortho-para transitions in this 
molecule, i.e., transitions caused by the magnetic-field gradient 
from a collision partner or a surface. There is room for the 
conversion by quantum relaxation only if direct transitions are 
unimportant. Below we describe the two ``direct'' contributions to 
the conversion which are produced by surface or bulk collisions.

\subsection{Direct conversion on surfaces}

For the case of ortho and para H$_2$, conversion on catalytic 
surfaces is by far the most efficient mechanism (see e.g. 
\cite{Ilisca92,Ilisca70PRL,Ilisca86PRL,Ilisca91PRL}). This may not 
be surprising in view of the extremely slow gas-phase conversion. 
But also for CH$_3$F, conversion on surfaces plays a role.
As was demonstrated in Refs.\cite{Chap85CP,Nagels98PhD}, CH$_3$F 
nuclear spin conversion can be quite different for different wall 
materials. The most rapid conversion (in fact instantaneous on the 
experimental time scale) was observed on Fe(OH)$_3$ powder 
\cite{Chap85CP}. The rate of conversion on this catalyst was limited, 
most likely, by diffusion. 

Glass surfaces appeared to be rather inert for the CH$_3$F conversion. 
The surface contribution to the conversion is given by a cutoff of 
the linear fit of the conversion rate vs. pressure. The data in 
Figure~\ref{pressure} determines the surface contribution on the 
order of 10$^{-4}$~s$^{-1}$. This yields a probability per collision 
for the molecule to convert  $\simeq$10$^{-9}$ 
\cite{Chap85CP,Nagels98PhD}. 
There are other surfaces which do not seem to produce rapid 
conversion, e.g., aluminum and gold. On the other hand, conversion 
on the surface of magnetic tape was found to be rather fast 
\cite{Nagels98PhD}.
 
\subsection{Direct conversion by collisions}

Another possible mechanism of direct ortho-para transitions is 
due to collisions in the bulk. In the case of hydrogen, it is 
known \cite{Farkas35} that oxygen molecules speed up hydrogen 
conversion dramatically, although in a time domain different from 
the CH$_3$F conversion. This phenomenon is well understood 
\cite{Wigner}. The conversion is due to the steep magnetic field 
gradient  produced by the collision partner. 

Conservation of the nuclear spin symmetry of symmetric tops in 
molecular collisions was confirmed in a number of studies using 
various experimental approaches
\cite{Oka73AAMP,Chesnokov77JETP,Harradine84JCP,Everitt89JCP,
Matsuo91JCP,Shin91JCP}. This is not surprising, because we know 
now that spin isomers are very stable species.  

Nevertheless, it would be useful to know the conversion rate in 
direct collisions. First of all we would like to stress that 
conservation of nuclear spin symmetry in molecular collisions is 
an approximate selection rule, although a very precise one. Like 
in the case of hydrogen, it can be violated if an inhomogeneous 
magnetic field is produced by a colliding molecule. From the 
measurements of the CH$_3$F conversion rate one can estimate an 
upper bound for the cross section of direct spin-changing collisions 
$\sigma(p|o)$. Using the formula for the gas-kinetic relaxation rate
$\gamma_{coll}=2N\sigma(p|o)v$ and the experimental value for the 
conversion rate 0.01~s$^{-1}$ \cite{Chap90JETP,Nagels96PRA} one 
obtains the estimation
\begin{equation}
     \sigma(p\mid o) \le 2.5\cdot10^{-24}~cm^2.
\label{sigma}
\end{equation}
Cross sections of this order are hardly accessible by conventional 
collision 
experiments, like molecular beams, or double resonance techniques.

In order to reveal the role of the magnetic moment of the collision 
partner, the $^{13}$CH$_3$F
spin conversion was studied in two buffer gases: N$_2$ and O$_2$ 
which have quite similar pressure broadening coefficients and, 
consequently, should have similar contributions to the conversion 
due to quantum relaxation \cite{Nagels96PRA}. On the other hand, 
the magnetic moments of N$_2$ and O$_2$ are different by three order 
of magnitude, which can result in extra contribution to the conversion 
in the case of oxygen.
From the data for $^{13}$CH$_3$F presented in Figure~\ref{buff}, 
one can conclude that the role of the oxygen magnetic moment is too 
small to be visible on the much larger background which is due to 
conversion by quantum relaxation. 
Thus in order to observe direct ortho-para transitions one has to 
suppress the contribution by quantum relaxation. It was proposed in 
\cite{Nagels98CPL} to search for spin-violating collisions by 
studying spin conversion in $^{12}$CH$_3$F, in which the conversion 
by quantum relaxation is slower by almost two orders of magnitude 
than in $^{13}$CH$_3$F.

The results of the measurements are presented in Figure~\ref{vio}. 
One can see that the 
$^{12}$CH$_3$F conversion is considerably faster in collisions with 
O$_2$ than with N$_2$. The difference in rates can be attributed 
\cite{Nagels98CPL} to  direct ortho-para transitions. The estimation 
for the cross section from the data in Figure~\ref{vio} reads
\begin{equation}
     \sigma(p|o) = (6\pm0.8)\cdot10^{-26}~cm^2.
\label{direct}
\end{equation}

We see from this result that even if CH$_3$F is diluted in 
paramagnetic oxygen, the direct ortho-para transitions are very 
improbable if compared with ordinary kinetic cross sections.
To understand to what extend the direct transition can influence the 
conversion in a ``nonmagnetic'' environment (like pure CH$_3$F), 
one needs to scale the cross section $\sigma(p|o)$ for the various 
magnetic moments. Similar to the case of hydrogen \cite{Wigner}, the 
cross section of direct transitions for CH$_3$F scales with $\mu^2$ 
where $\mu$ is the magnetic moment of the collision partner 
\cite{Chap96CPL}. Consequently, direct transitions can be neglected 
for CH$_3$F spin conversion in a nonmagnetic environment.

\section{CONCLUSION}
 It is well established by now that the nuclear spin isomers of 
 molecules can be separated in the gas phase by using the 
 Light-Induced Drift technique. The enriched gas samples can be 
 stored in a separate test volume for further investigations of 
 their properties.

In the present review we have used CH$_3$F to investigate the 
mechanism behind nuclear-spin conversion. This molecule was found 
to have conversion rates in a convenient time domain, and turned 
out to be very versatile for investigating various aspects of the 
conversion process.
 The conversion was studied at various pressures, temperatures and 
 gas compositions and for two isotopic species. The experimental 
 results obtained for the CH$_3$F spin isomers revealed that the 
 conversion in the bulk is governed by a specific process which can 
 be called {\underline{quantum relaxation}}. This mechanism is based 
 on the {\underline{intramolecular}} mixing of molecular quantum 
 states and {\underline{collisional}} destruction of the quantum 
 coherence between the mixed states. 

Various key points in the CH$_3$F conversion by quantum relaxation 
were clarified. It includes, first of all, the knowledge of the most 
important levels in CH$_3$F which contribute substantially to the 
conversion through mixing by intramolecular perturbations. Second, 
the mixing mechanism in $^{13}$CH$_3$F is shown to be the spin-spin 
interaction between the molecular nuclei. For $^{12}$CH$_3$F, on the 
other hand, there is evidence that the spin-rotation perturbation is 
the most important while spin-spin interactions play a minor role. 
Third, the decay rate of the coherence between ortho and para states 
by collisions in the bulk was determined experimentally.
All  these data gave an internally consistent picture of the spin 
conversion in CH$_3$F.

The mechanism responsible for nuclear spin conversion in CH$_3$F was 
tested experimentally in a few ways. The most decisive experiments 
are the observation of level-crossing resonances in the conversion 
rate and the observation of the slowing down of the conversion with 
increasing gas pressure (``Quantum Zeno Effect'').

Prospects for further development in the field of spin isomers look 
optimistic. Even if one restrict oneself to the spin isomers separation 
method based on  LID, there is a large variety of investigations possible. 
The LID effect already proved to be an efficient separation tool for 
a number of molecules like NH$_3$, C$_2$H$_4$, H$_2$O which all have 
nuclear spin isomers. There is no doubt that investigations of 
molecular  spin isomers similar to that for CH$_3$F can be performed 
with various other molecules. 

\section{ACKNOWLEDGMENTS}

The results presented in this review were obtained in a long term 
project. We are indebted to many colleagues for fruitful and 
pleasant cooperation. More specifically, we would like to name the 
following colleagues (in the order in which they joined the project) 
V.N.~Panfilov, V.P.~Strunin, L.N.~Krasnoperov, A.E.~Bakarev, 
D.~Papou\v{s}ek, J~Demaison, B.~Nagels, M.~Schuurman, D.A.~Roozemond, 
N.~Calas, P.~Bakker, E.~Ilisca, M.~Irac-Astaud, K.~Bahloul. 

This project was made possible by financial support from the 
Netherlands Organization for 
Scientific Research (NWO), the Russian Academy of Sciences and the 
RFBR grant 98-03-33124a.

\newpage

TABLE 1. Close ortho-para level pairs in CH$_{3}$F and calculated 
conversion rates ($\gamma$) induced by spin-spin ($SS$) and spin 
rotation ($SR$) interactions.

\begin{tabular}{ccccc}
\hline\hline
    & level pair & $\omega_{\alpha'\alpha}/2\pi^{(1)}$ & $\gamma _{SS}/P$ & 
                                                  $\gamma _{SR}/P$ \\ 
Molecule & $J',K'$-$J,K$& (MHz) & (10$^{-3}$ s$^{-1}$/torr) & 
(10$^{-3}$ s$^{-1}$/torr) \\ \hline
$^{13}$CH$_{3}$F & 11,1-9,3  &  130.99$\pm $0.15  & 7.7 & -- \\ 
                 & 21,1-20,3 & -351.01$\pm $0.16  & 4.4 & 1.7$^{(2)}$
 \\ \hline
$^{12}$CH$_{3}$F & 9,2-10,0  & 8591.7$\pm$0.3 & 9.3$\cdot10^{-3}$ & 
1.3$\cdot10^{-3}$ \\ 
                 & 15,7-17,6 & 1745.7$\pm$2.3 & 6.8$\cdot10^{-3}$ &  -- \\ 
                 & 28,5-27,6 & 1189.2$\pm$1.5 & 16$\cdot10^{-3}$  & 
                 $\simeq0.3^{(3)}$\\ 
                 & 51,4-50,6 &  -41.3$\pm$2.5 & 28$\cdot10^{-3}$  & 
                 0.1 \\ \hline
\end{tabular}

\vspace{1cm}

$^{(1)}$Molecular parameters for $^{13}$CH$_{3}$F from \cite{Papousek94JMS}
and for $^{12}$CH$_{3}$F from \cite{Papousek93JMS33}.

$^{(2)}$Calculated assuming spin-rotational tensor component 
C$_{22}$=1~kHz \cite{Guskov95JETP}.

$^{(3)}$Calculated assuming spin-rotational tensor component 
C$_{21}$=3.6~kHz and a value for $\Gamma=2\cdot10^7$~s$^{-1}$/torr
\cite{Ilisca98PRA}.

\newpage

%\vspace*{2cm}

\begin{figure}[htb]
\centerline{\psfig
{figure=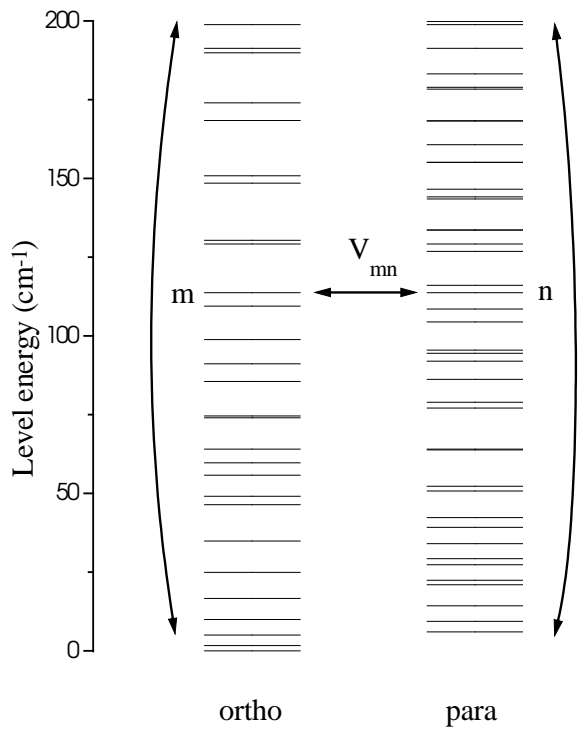,height=12cm}}
\caption{\label{subs} The rotational states of $^{13}$CH$_{3}$F below 
200~cm$^{-1}$ in the ground vibrational state  calculated using the 
molecular parameters from \cite{Papousek94JMS}. The level pair 
($J'$=11,$K'$=1)--($J$=9,$K$=3) is shown to be mixed by intramolecular 
perturbation $\hat V$. The bent lines indicate collisional transitions 
inside the ortho and para subspaces.}
\end{figure}

\newpage

\vspace*{1cm}

\begin{figure}[h]
\centerline{\psfig
{figure=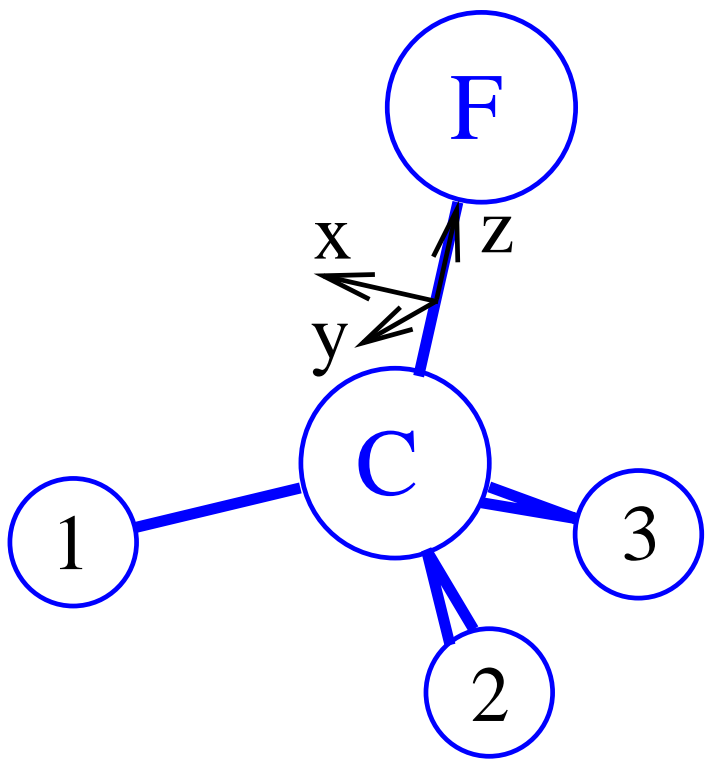,height=10cm}}
\caption{\label{coord} Numbering of the hydrogen nuclei in CH$_3$F 
and definition of the molecular coordinate system.}
\end{figure}

\newpage

\vspace*{1cm}

\begin{figure}[h]
\centerline{\psfig
{figure=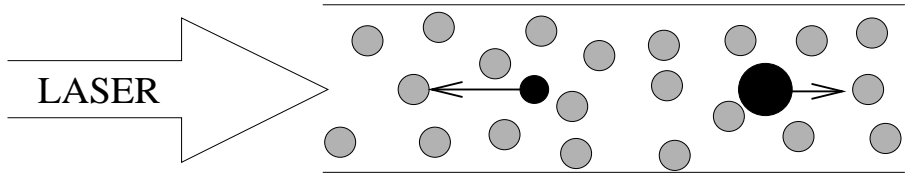,width=12cm}}
\caption{\label{lid} The principle of Light-Induced Drift (LID) 
\cite{Gel79JETPL}. Due to velocity-selective excitation combined 
with a state-dependent kinetic cross section, the light-absorbing 
species (black circles) exhibits anisotropic diffusion through the 
optically inert species (gray circles). Here we have assumed laser 
detuning in the blue wing ($\Omega>0$) and increased transport 
collision rate upon excitation, resulting in Light-Induced Drift 
of the absorbing species towards the laser.}
\end{figure}

\newpage

%\vspace*{1cm}

\begin{figure}[htb]
\centerline{\psfig
{figure=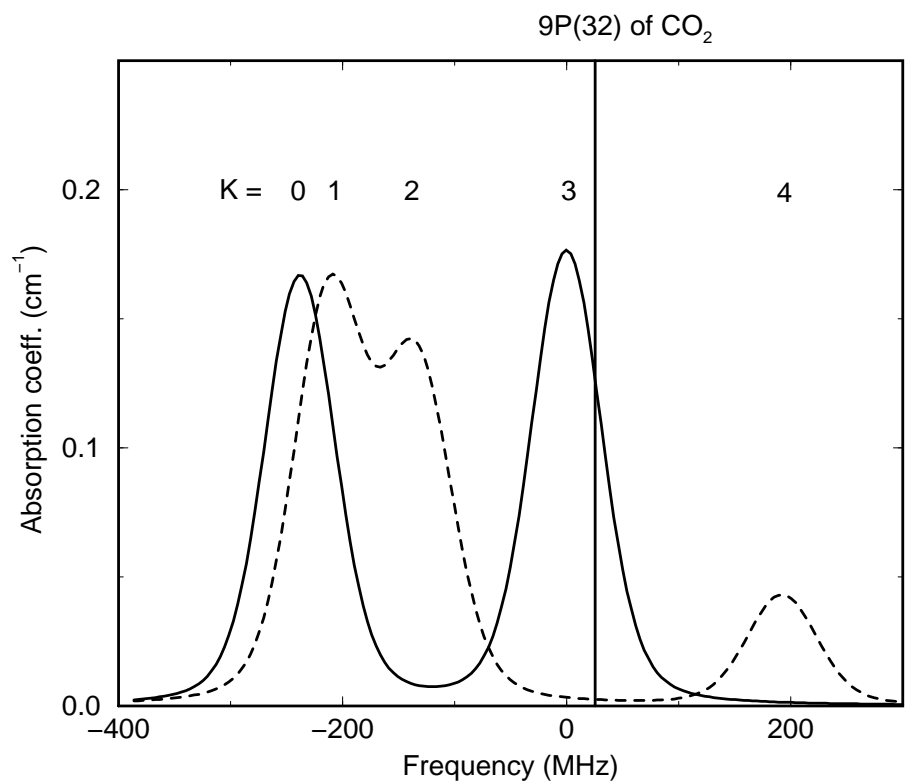,width=14cm}}
\caption{\label{spectr}The $R(4,K)$ absorption lines of ortho (--) 
and para (-\,-) $^{13}$CH$_3$F in the vicinity of the P(32) line of 
a CO$_2$ laser in the 9.6~$\mu$m band. The spectrum was calculated 
for pressure $P$=0.5~torr of $^{13}$CH$_3$F  using the molecular 
parameters from \cite{Papousek94JMS,Freund74JMS}.} 
\end{figure}

\newpage

\vspace*{1cm}

\begin{figure}[h]
\centerline{\psfig
{figure=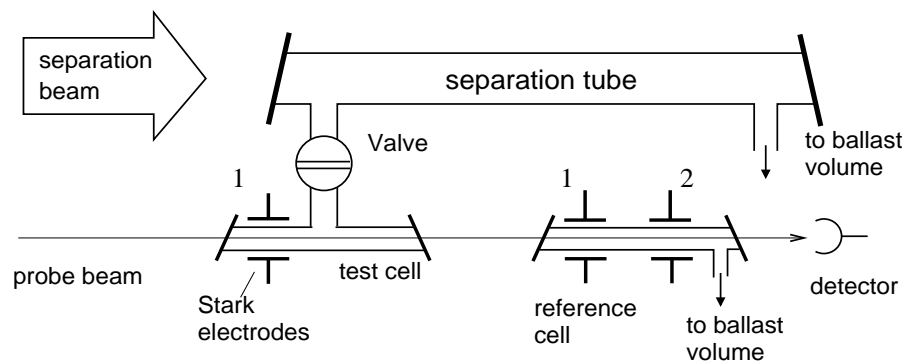,width=12cm}}
\caption{\label{novosib} Schematic of the Novosibirsk setup 
\cite{Chap90JETP}. Separation is achieved in the long upper tube. 
For increased sensitivity and long-term stability in detection, this 
differential method makes use of two detection cells with phase-shifted 
Stark modulation by electrodes 1 and an additional Stark modulation by 
electrodes 2 (see text).}
\end{figure}

\newpage

\vspace*{1cm}

\begin{figure}[h]
\centerline{\psfig
{figure=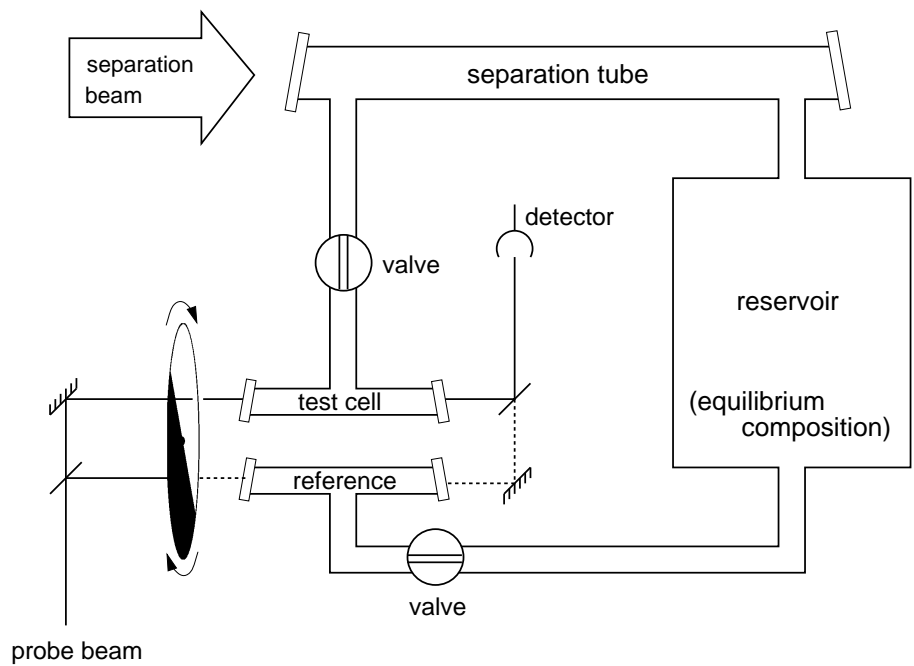,width=12cm}}
\caption{\label{leiden} Schematic of the Leiden setup \cite{Nagels96PRA}. 
After separation by LID, an enriched and a reference sample are 
isolated by closing two valves, and the conversion process is probed 
by absorption. For conversion measurements at elevated temperatures a 
two-compartment oven was connected to the test and reference cells 
\cite{Nagels98PRA}. For conversion measurements in an electric field 
a Stark cell was added to the test cell \cite{Nagels96PRL}.}
\end{figure}

\newpage

\vspace*{1cm}

\begin{figure}[h]
\centerline{\psfig
{figure=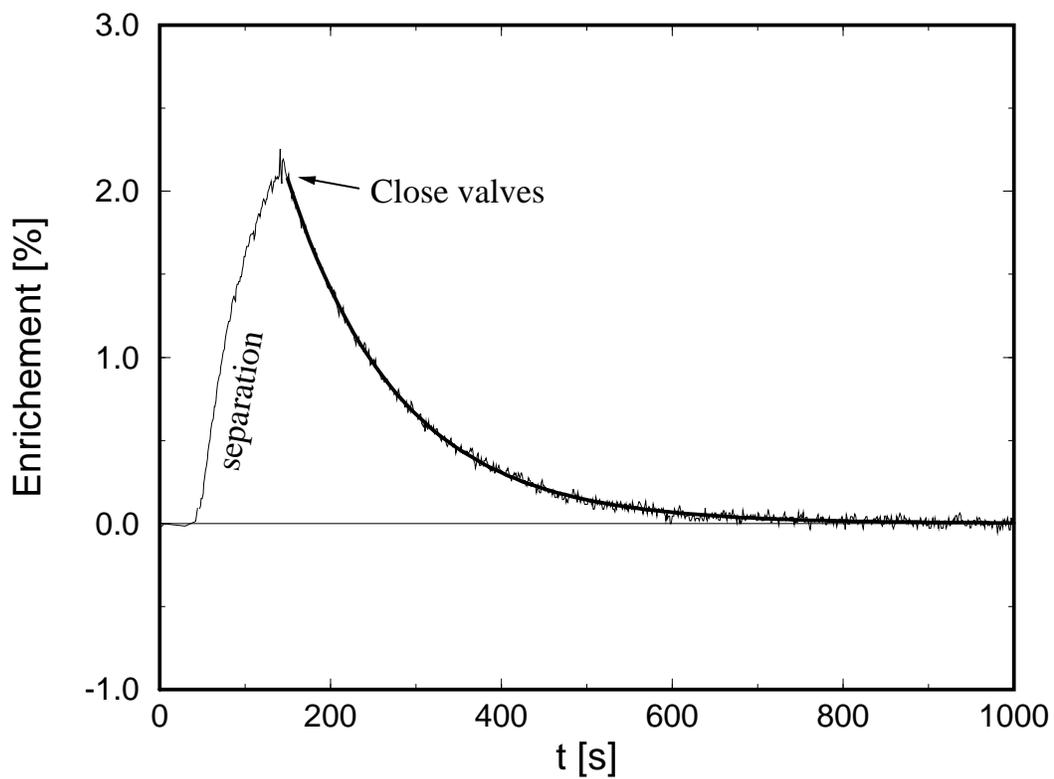,height=12cm}}
\caption{\label{decay} Typical enrichment decay curve as measured  
by absorption, for the $^{13}$CH$_3$F spin conversion at pressure 
0.6~torr. The smooth curve is an exponential fit to the data 
\cite{Nagels96PRA}.}
\end{figure}

\newpage

\vspace*{1cm}

\begin{figure}[h]
\centerline{\psfig
{figure=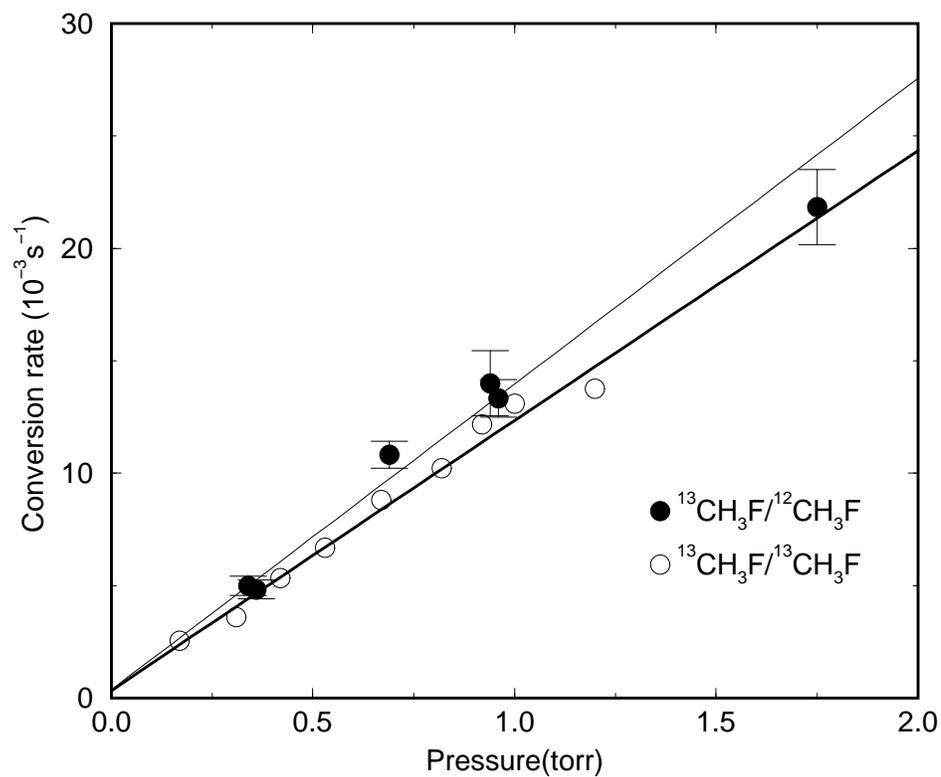,height=12cm}}
\caption{\label{pressure} The pressure dependence of the 
$^{13}$CH$_3$F conversion rate   as a function of pressure. 
($\bullet$) -- $^{13}$CH$_3$F
in $^{12}$CH$_3$F as a buffer gas \cite{Chap90JETP}; 
($\circ$) -- pure $^{13}$CH$_3$F \cite{Nagels96PRA}. The solid 
lines give a linear fit for the two data sets. }
\end{figure}

\newpage

\vspace*{1cm}

\begin{figure}[h]
\centerline{\psfig
{figure=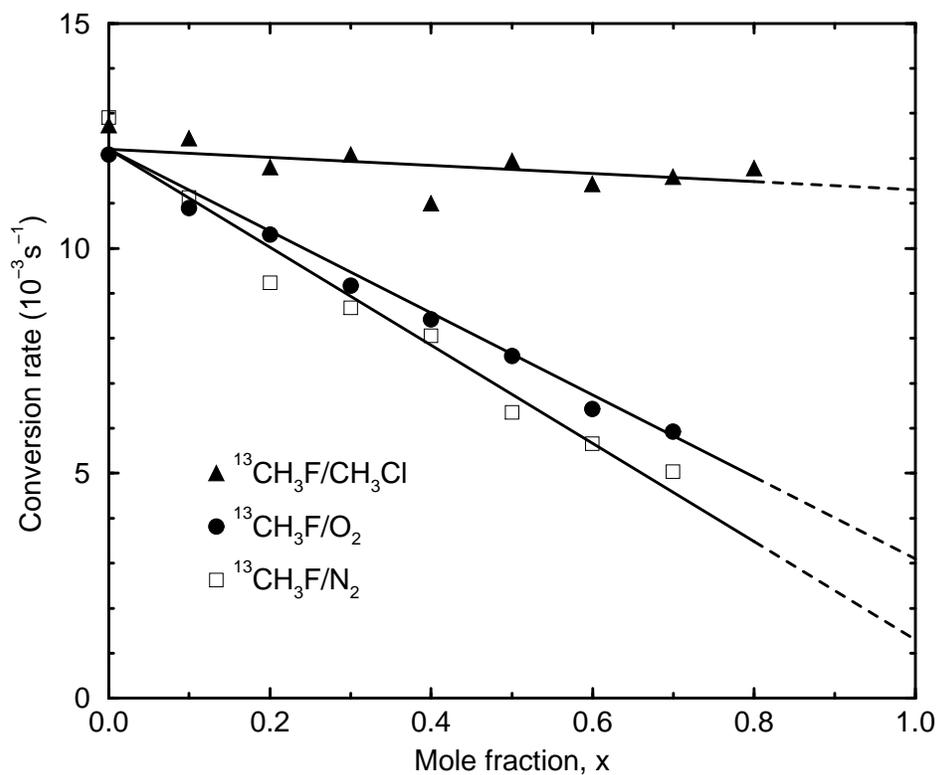,height=12cm}}
\caption{\label{buff} The $^{13}$CH$_3$F conversion rate in binary 
mixtures as a function of mole fraction of CH$_3$Cl, O$_2$, or N$_2$. 
The total pressure $P$=1~torr, $T=297$~K. Extrapolation to $x=1$ yields 
the conversion rate in collisions with the corresponding buffer gas 
\cite{Nagels96PRA}.}
\end{figure}

\newpage

\vspace*{1cm}

\begin{figure}[h]
\centerline{\psfig
{figure=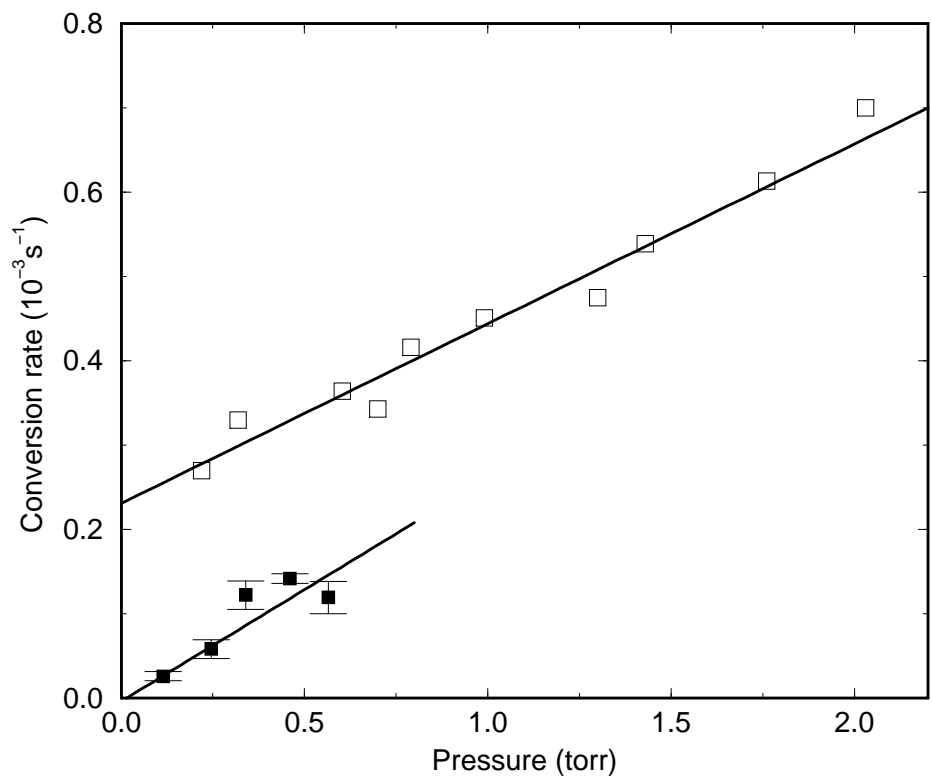,height=12cm}}
\caption{\label{isotope} The pressure dependence of the conversion 
rate in $^{12}$CH$_3$F. Data marked by open squares are from 
\cite{Nagels98PRA}; filled squares are from \cite{Chap90JETP}.}
\end{figure}

\newpage

\vspace*{1cm}

\begin{figure}[h]
\centerline{\psfig
{figure=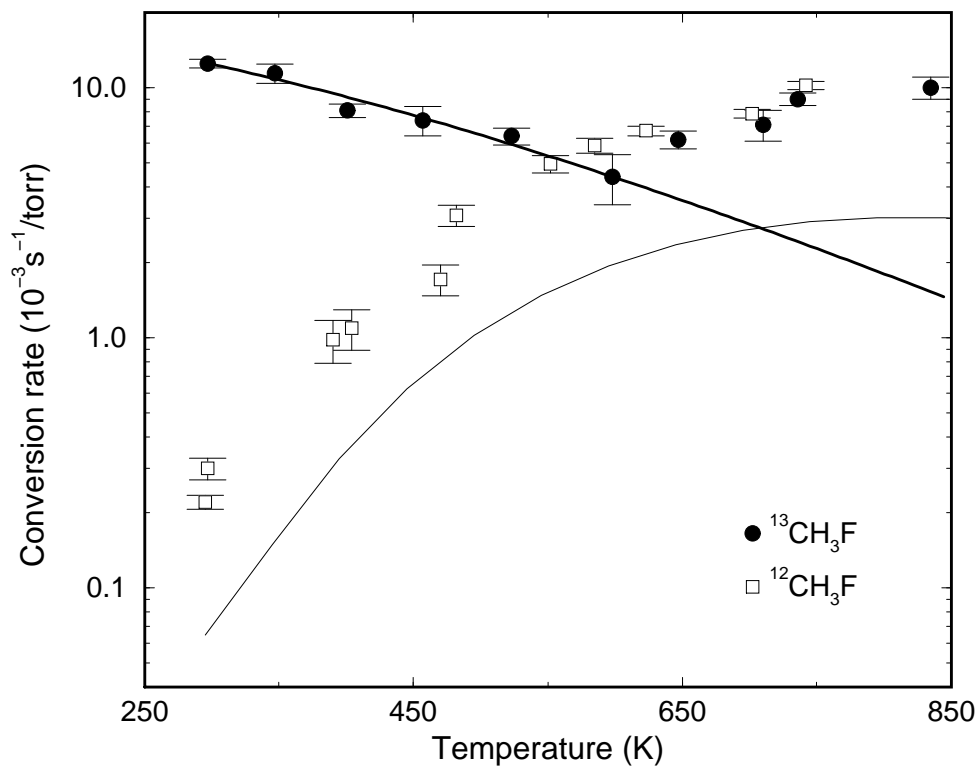,height=12cm}}
\caption{\label{temp} The conversion rates in CH$_3$F at various 
temperatures \cite{Nagels98PRA} (revised). The solid lines represent 
the calculated conversion rates in $^{13}$CH$_3$F and $^{12}$CH$_3$F 
assuming the same $\Gamma(T)$ for both molecules (see Eq.~(\ref{gT})) 
and  mixing of ortho and para states only by spin-spin interaction. }
\end{figure}

\newpage

%\vspace*{1cm}

\begin{figure}[htb]
\centerline{\psfig
{figure=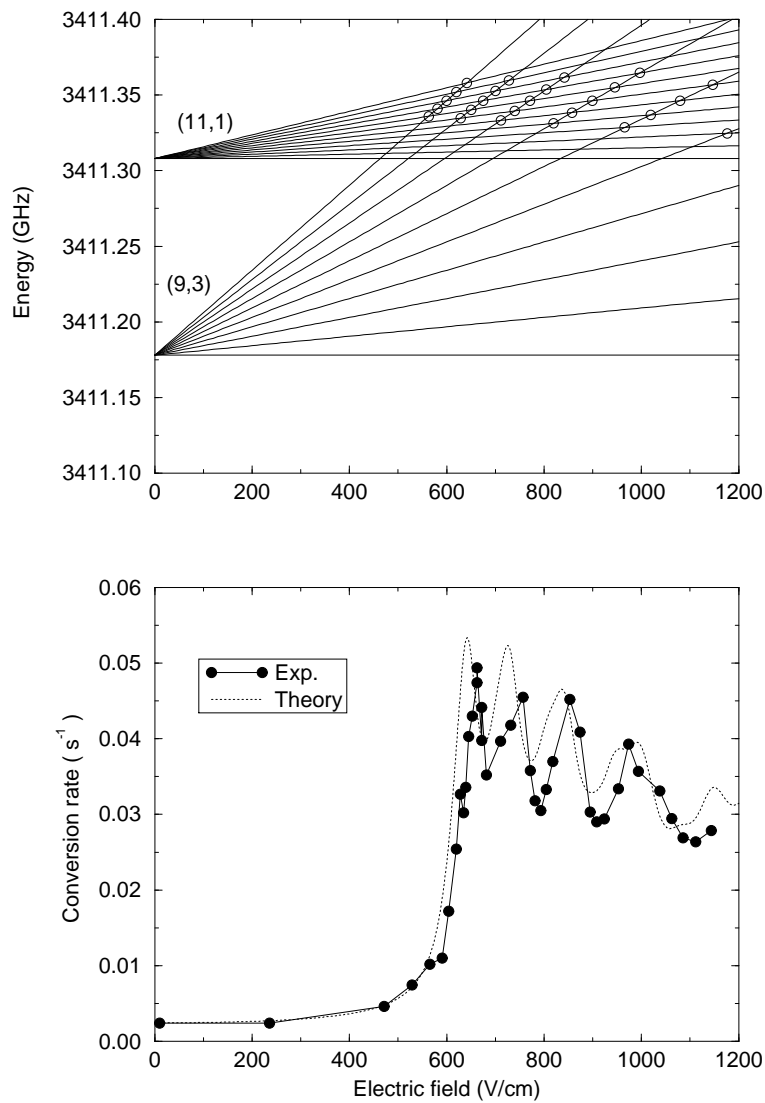,height=16cm}}
\caption{\label{stexp} Upper panel: Splitting of the (11,1) and 
(9,3) levels by an electric  field. Crossings which should contribute 
to the conversion through mixing by spin-spin interaction 
($|\Delta M|\le2$) 
are marked. Lower panel: Experimental and theoretical spin conversion 
rates  $\gamma({\cal E})$ in $^{13}$CH$_3$F as a function of the  
electric field strength. Gas pressure is
0.20~torr (26.6 Pa). The experimental points are connected to guide 
the eye \cite{Nagels96PRL}.}
\end{figure}

\newpage

%\vspace*{1cm}

\begin{figure}[htb]
\centerline{\psfig
{figure=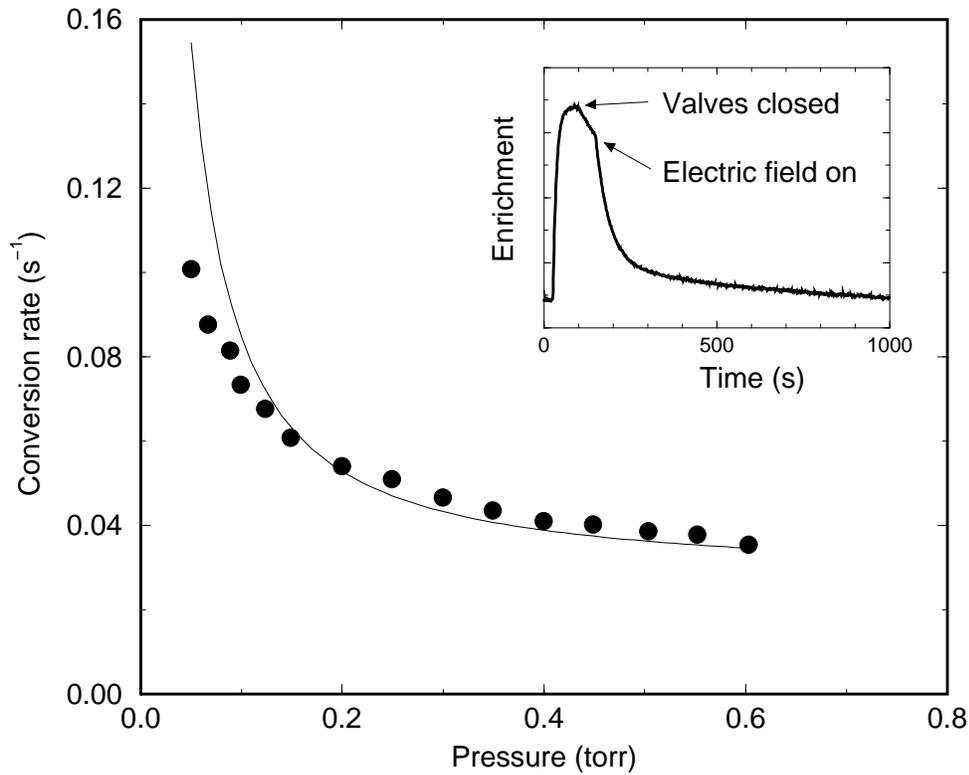,height=12cm}}
\caption{\label{zeno} The ortho--para conversion rate in $^{13}$CH$_3$F 
as a function of pressure in a static electric field of 652.8~V/cm : 
experiment ($\bullet$) and theory (--). The field is chosen such that 
the $M'$=11 and $M$=9 magnetic sublevels of the levels ($J'$=11, $K'$=1) 
and ($J$=9, $K$=3) become degenerate (see Figure~\ref{stexp}). The inset 
shows a typical decay curve \cite{Nagels97PRL}.}
\end{figure}

\begin{figure}[h]
\centerline{\psfig
{figure=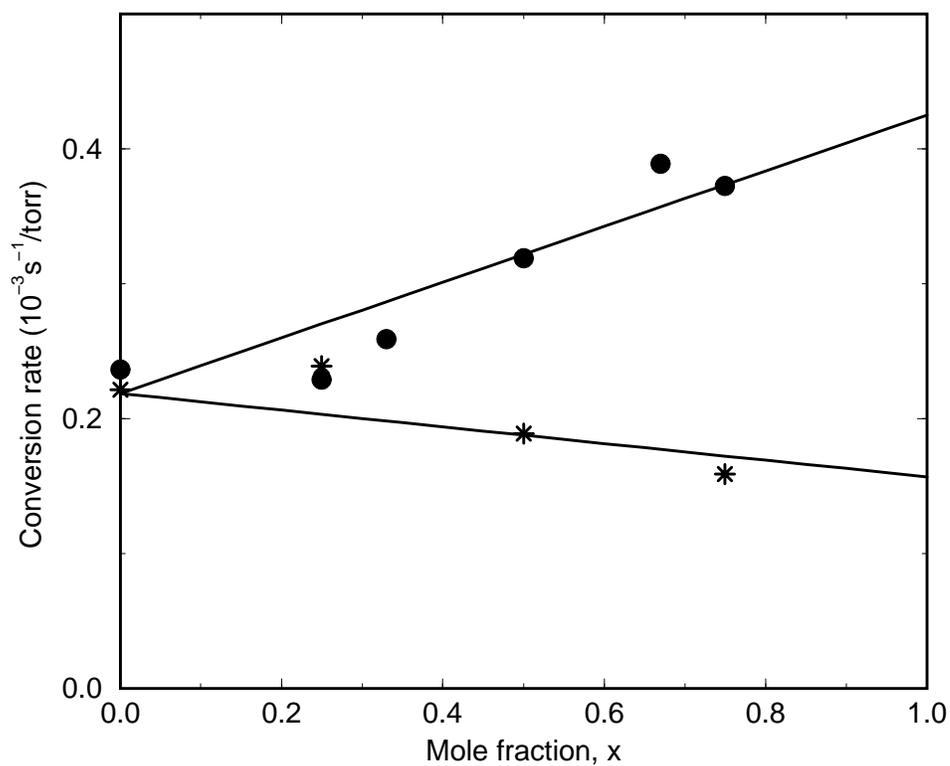,height=12cm}}
%\vspace*{0.5cm}
\caption{\label{vio} The experimental data for the nuclear spin 
conversion in $^{12}$CH$_3$F molecules as a function of mole fraction 
of nitrogen ($\ast$), and oxygen ($\bullet$) \cite{Nagels98CPL}. The 
total pressure is 1~torr. Note that the conversion rate 
{\underline{increases}} with increasing O$_2$ mole fraction, in 
contrast to the behavior of $^{13}$CH$_3$F displayed in Figure~\ref{buff}.}
\end{figure}

\end{document}